\documentclass[fleqn,usenatbib]{mnras}

\usepackage{newtxtext,newtxmath}

\usepackage[T1]{fontenc}
\usepackage{ae,aecompl}
\usepackage{multicol}
\usepackage{titlesec}

\usepackage{graphicx}	
\usepackage{amsmath}	
\usepackage{appendix} 
\usepackage{threeparttable}

\def\HII{{H{\sc ii}}}

\title[Sulphur as abundance tracer in galaxies]{On the use of Sulphur as a tracer for abundances in galaxies }

\author[A. I. Díaz and S. Zamora]{Ángeles I. Díaz,$^{1,2}$\thanks{E-mail: angeles.diaz@uam.es}
S. Zamora $^{1,2}$
\\
$^{1}$Department of Theoretical Physics, Universidad Autónoma de Madrid, Spain \\
$^{2}$CIAFF, Universidad Autónoma de Madrid, Spain \\
}

\date{Accepted XXX. Received YYY; in original form ZZZ}

\pubyear{2021}

\begin{document}
\label{firstpage}
\pagerange{\pageref{firstpage}--\pageref{lastpage}}
\maketitle

\begin{abstract}
We present a methodology for the use of sulphur as global metallicity tracer in galaxies, allowing performing a complete abundance analysis using only the red-to-near infrared spectral region. We have applied it to a compilation of high-quality data split into two samples:  \HII\ regions (DHR) in spiral and irregular galaxies, and dwarf galaxies dominated by a strong starburst (HIIGal).
Sulphur abundances have been derived by direct methods under the assumption of an ionisation structure composed of two zones: an intermediate one where S$^{++}$ is originated and a low ionisation one where S$^+$ is formed. 
Ionisation correction factors (ICF) have been calculated from the Ar$^{2+}$/Ar$^{3+}$ ratio and are shown to correlate with the hardness of the radiation field. Only about 10\% of the objects show S$^{3+}$ contributions to the total abundance larger than 30\%.
A good correlation exists between sulphur abundance and ionising temperature with low metallicity objects being ionised by hotter stars. No correlation is found between ionisation parameter and total S/H abundance.
Most of the HIIGal objects show S/O ratios below the solar value and a trend for increasing S/O ratios with increasing sulphur abundances while DHR objects show S/O ratios larger than solar and a tendency for lower S/O ratios for higher metallicities. Finally, we present a calibration of the sulphur abundance through the S$_{23}$ parameter that remains single valued up to sulphur abundances well beyond the solar value. S$_{23}$ is independent of the ionisation parameter and only weakly dependent on ionising temperature. 

\end{abstract}




\section{Introduction}


Metals -- any chemical element heavier than helium -- are formed inside stars. Therefore, it is expected that metal enrichment in the universe started soon after the formation of the first short-lived massive stars that promptly returned to the interstellar medium (ISM), in explosive events, the newly synthesised chemical elements. It is thought that these ejecta, at some point, mixed with the surrounding gas helping to its cooling and propitiated the appearance of a new generation of stars, hence giving rise to the cycle of cosmic chemical evolution. 

Metals play a very important role in star formation and stellar evolution. Amongst other things they control the cooling of the interstellar gas, thus allowing the formation of stars; they affect the radiation transport, through the opacities involved in the different microscopic processes; and they have the most important role in the dust formation and mass loss from stars. Different chemical element abundances can be derived from the analysis of the ionised nebulae (\HII\ regions) that accompany the young, massive stellar component in galaxies. These massive stars illuminate and ionise cubic kilo parsec sized volumes of the ISM and hence are visible in far away galaxies. \HII\ regions emit strong hydrogen and helium recombination lines and forbidden lines from a variety of heavy elements: O, Ne, N, S, Ar, etc., and, with appropriate constraints on physical conditions such as the electron temperature and density of the emitting gas, it is possible to derive ionic and total elemental abundances and thus obtain the composition of the ISM.

Recombination lines yield the most accurate abundances because of their weak dependence on the nebular electron temperature (in fact, helium abundances can be determined with an accuracy higher than 5 \%). But, unfortunately, most of the observed emission lines in nebulae are collisionally excited and their intensities depend exponentially on electron temperature which, in principle, can be derived from appropriate line ratios of nebular to auroral lines, the latter one being intrinsically weak and difficult to detect. However, it is possible to use the cooling properties of the ionised gas in order to produce empirical calibrations relating the emission line intensity ratios of nebular lines, much stronger and easily observable, with the abundance of a given element.

Over many years oxygen has been used as the main present abundance tracer in galaxies with star formation. In fact, it is the oxygen abundance which is referred to when the term "metallicity" is employed in this context. Oxygen, the most abundant element after hydrogen and helium in \HII\ regions, is produced in high mass stars (m > 8 M$_{\sun}$) and ejected in supernova explosions. Forbidden collisionally excited emission lines of singly and doubly ionised oxygen are observed in the blue-yellow region of the spectrum ([OII] $\lambda\lambda$ 3727,29 \AA\ and [OIII] $\lambda\lambda$ 4959, 5007 \AA ) accessible with conventional spectrographs from many decades back and a well developed methodology to compute their abundances exists  \cite[see, for example,][]{1969BOTT....5....3P}, although this is not exempt of problems, as we will see later.

On the other hand, the use of sulphur as an abundance tracer has frequently been  overlooked. Although less abundant than oxygen, sulphur is also produced in massive stars and therefore its yield is expected to follow closely that of oxygen. Nebular S/H abundances are therefore expected to also follow those of O/H and hence the S/O ratio should remain constant at about the solar value, log S/O $\simeq$ -1.7 \citep{2009ARA&A..47..481A}. Empirical tests to validate this have been performed apparently confirming the cosmic nucleosynthetic ratio \cite[see][for a recent relation between S/O and O/H]{2020ApJ...893...96B}, although the results of some works suggest that this relation should be explored further, particularly at the not well sampled metallicity ends: extremely metal deficient \HII\ galaxies (i.e. very low O/H) and \HII\ regions in the inner zones of galaxy discs (i.e. metal rich central zones with the highest O/H abundances). It should be taken into account that, although a certain amount of oxygen in diffused clouds is depleted, this is not the case for sulphur which is almost undepleted in the low density, ISM \citep[][]{2009ApJ...700.1299J,2021A&A...648A.120R}.


The nebular emission lines of sulphur are analogous to those of oxygen, with the singly and doubly ionised sulphur lines occurring at longer wavelengths:  [SII] at $\lambda\lambda$ 6716,31 \AA\ and [SIII] at $\lambda\lambda$ 9069, 9532 \AA , and, in principle, the same methodology can be used for the direct derivation of its abundance, but presenting several advantages. On the theoretical side, although their intensities depend exponentially on electron temperature, as in the case of oxygen,  this dependence is lower due to the lower energies involved in the atomic transitions and this fact renders the lines observable even at solar or over-solar abundances \citep{2007MNRAS.382..251D}. This also applies to the [SIII] auroral line, at $\lambda$ 6312 \AA\ which can be detected and measured at, at least, solar abundances \citep{1994A&A...282L..37K}. Furthermore, the region in the ionised nebula where  $S^{2+}$ originates  practically covers the $O^{+}$ and $O^{++}$ ones \citep[see][]{1992AJ....103.1330G} and hence the corresponding [SIII] electron temperature, T$_e$([SIII]), can be taken as representative of the whole nebula. On the observational side, the fact that the [SIII] $\lambda$ 6312 \AA\  falls in a spectral region of high instrumental sensitivity and is weakly affected by underlying stellar population absorption, makes it possible to derive $S/H$ by direct methods up to solar values. Also, the lines of both [SII] and [SIII] can be measured relative to nearby hydrogen recombination lines (H$\alpha$, P$_9$, P$\epsilon$), thus minimising the effect of any reddening and/or calibration uncertainties. On the negative side, the strong nebular [SIII] lines shift out of the far red spectral region for redshifts higher than 0.1. However this drawback is being overcome by the existence of new, efficient, infrared spectrographs. For instance, the XShooter spectrograph can accommodate in the visible and the infrared J, H and K bands the emission lines needed to apply this methodology minimizing sky subtraction problems for more than 106 SDSS galaxies in different redshift ranges between 0.17 and 1.5 and presumably a good portion of these should present emission lines associated to present star formation. Observations made with NIRSpec on board  the JWST, whose spectroscopic range starts at 6000 Å, would be able to provide data for objects with redshifts between 0 and 4.24.

Despite the mentioned advantages, apart from the works of \citet{2000MNRAS.312..130D,2005MNRAS.361.1063P,2006A&A...457..477K,2007MNRAS.382..251D}, very little attention has been given to the [SIII] lines in the literature. \citet{2002ApJS..142...35K} mentioned briefly the [SIII]/[SII] ratio for the estimation of the ionisation parameter suggesting the existence of errors in its derivation. They also mentioned the possibility of estimating abundances using diagnostics involving the strong sulphur lines and concluding that: "this diagnostic does not reliably estimate abundances for any metallicity range, and should therefore not be used".
Contrary to this, later work by \citet{2011MNRAS.415.3616D} concluded that the use the [SIII]/[SII] vs [[OIII]/[NII] diagnostic gives O/H values nearest to those derived from direct electron temperature methods and similar to those resulting from the application of detailed modelling.


At any rate, as expected from the exponential dependence on temperature of the intensities of  collisionally excited lines, when the cooling is dominated by metals the electron temperature depends inversely on their abundance and the auroral lines become too faint to be detected and measured. But it is possible to establish a relation between increasing metal abundance (i.e. decreasing electron temperature) and decreasing intensity of the strong nebular lines.  Following this general idea, a plethora of so called “empirical calibrations” have been devised, starting with the ratio of [OIII] $\lambda$ 5007 \AA /H$\beta$ \citep{1976ApJ...209..748J}, the ratio of [OIII] $\lambda$ 5007 \AA /[NII] $\lambda$ 6584 \AA \citep{1979A&A....78..200A} and the ratio of ([OII]$ \lambda\lambda$ 3727,29 \AA + [OIII] $\lambda\lambda$ 4959,5007 \AA )/H$\beta$ \citep{1979MNRAS.189...95P}, commonly referred to as R$_{23}$. The fact that this latter index is not affected by variations in relative abundances, as is the case of [OIII]/[NII], and remains essentially constant within a given giant \HII\ region despite variations in excitation \citep{1987MNRAS.226...19D}, has made of it the most widely used abundance calibrator. The calibration is, however, far from being ideal due to its intrinsic two-folded nature, since, at the highest metallicities, the high efficiency of the oxygen as a cooling agent {\em decreases} the strength of the oxygen emission lines, while, at low metallicities, the cooling is mainly exerted by hydrogen recombination lines and the oxygen line strengths {\em increase} with metallicity. A value of of O/H $\simeq$ 1.3 $\times$ 10$^{-4}$ (12+log(O/H) $\simeq$ 8.1) divides the two different abundance regimes. Although some line ratios like, for instance [NII]/H$\alpha$  have been proposed in order to break this degeneracy, the fact that a large number (up to 80 \% ) of \HII\ regions and star forming galaxies lie right on the turnover region is of great concern \citep[see e.g.][]{2009MNRAS.398..949P}.

In this respect, the use of the sulphur lines allows to overcome this caveat. Due to their nature, similar to that of the oxygen lines, the same line of reasoning may be put forward regarding the use of an S$_{23}$ parameter defined as ([SII] $\lambda\lambda$ 6716,31 \AA + [SIII] $\lambda\lambda$ 9069, 9532 \AA )/H$\beta$, as an alternative abundance indicator, which in fact presents several advantages against R$_{23}$: (1) due to the longer wavelengths of the lines involved, its relevance as a cooling agent starts at lower temperatures (higher metallicities) which makes the relation to remain single-valued up to solar abundances, at least; (2) their lower dependence on electron temperature, renders the lines observable even at over-solar abundances; and (3), as stated above, the lines of both [SII] and [SIII] can be measured relative to nearby hydrogen recombination lines thus minimising the effect of reddening and/or calibration uncertainties. For calibration purposes in this wavelength range, the sulphur lines can be easily scaled to the closest hydrogen recombination lines like:

\[S_{23}  = \frac{I([SII] 6716+6731)}{H\alpha} \cdot \frac{H\alpha}{H\beta}+\frac{I([SIII] 9069+9532)}{P_9} \cdot \frac{P_9}{H\beta}  \]

\noindent where H$\alpha$/H$\beta$ and P$_9$/H$\beta$ are the theoretical case B recombination ratios.

Given these properties, here we put forward a methodology for the use of spectroscopy in the red-to-near infrared wavelength range in order to derive physical properties and abundances of ionised regions, including the red lines from [SIII] $\lambda$ 6312 \AA\ and [SII] $\lambda$ 6717,31 \AA\ in the red up to  the far-red lines [SIII] $\lambda$ 9069,9532 \AA . These lines can a play a convenient analogue, in this wavelength range, of the role played by the [OII] and [OIII] lines in the optical. Among the advantages provided by this observational set up is the much lower sensitivity to reddening effects, something of especial relevance for the study of heavily reddened star-forming objects,  \HII\ regions in the Milky Way through the Galactic disc or inner regions of galaxies. Besides, most of the spectrographs being built for large telescopes at present are optimised for the red wavelength range (e.g. SDSS, MUSE) starting at about 4000-4800 \AA\ which makes worthy the effort to develop methodologies adequate to this new situation.
 
In the next section we present the sample of objects that we have used here to establish a good quality sulphur abundance calibration.  Section 3 is devoted to the direct derivation of sulphur and oxygen abundances, including the treatment of the ionisation correction factor (ICF), not yet well established, using selected data on \HII\ regions and \HII\ galaxies with measurements of the [SIV] infrared line fluxes, together with the predictions of photoionisation models. In Section 4 we present the proposed empirical calibration of S/H versus S$_{23}$. Section 5 is devoted to the discussion of the results obtained in this work and, finally, section 6 summarizes our main conclusions.

\section{Sample selection}

Data obtained using CCD detectors sensitive in the red to near-infrared spectral range started to appear in the mid 80's. First CCDs showed a great amount of fringing that was difficult to eliminate and required long flat field exposures. Their size was small and therefore the spectral coverage was also small. Several different exposures were required in order to cover a spectral range that would be sufficient to allow an adequate abundance analysis. Flux calibration was also difficult due to the lack of calibration star data up to 1 $\mu$m \citep{1985MNRAS.212..737D,1987MNRAS.226...19D}. The night sky lines, mainly O$_2$ and OH lines, are very intense in the far-red spectral region \citep{1992PASP..104...76O}, and therefore a good sky subtraction was crucial. This subtraction is better if moderate to high spectral resolution (600 mm $^{-1}$, R $\simeq$ 6000) is used \citep{2003A&A...407.1157H}. Also, atmospheric water absorption bands can show up very prominently. They depend on the level of humidity and can vary through the observing night. They can be eliminated by observing a star of almost featureless spectrum before and after the exposure at the same resolution \citep{1994MNRAS.267..904S,2015A&A...576A..77S}. Nowadays, most of these problems have been solved and very good quality spectra are routinely obtained \citep[see, for example,][]{2008MNRAS.383..209H}.

The general sample selected for this study consists of nebulae ionised by young massive stars residing in galaxies with different properties and they have been split into two. The first sub-sample labelled "DHR" comprises observations of diffuse Giant Extragalactic \HII\ Regions over spiral galaxy discs, including our own, and the Magellanic Clouds. It lists 256 independent observations. The second one, labelled \HII{Gal}, comprises compact galaxies whose spectra are entirely dominated by a current star formation burst and therefore show very intense emission line spectra with large equivalent widths and weak underlying stellar continua. This second sub-sample lists 95 independent observations and includes the galaxies with the lowest gas metallicities found.

Data in all the objects of the general sample have been taken from the literature and some of them correspond to very well studied objects (e.g. Orion nebula in the Milky Way Galaxy or 30Dor in the Large Magellanic Cloud) which have more than one set of data. They cover, at least, the spectral range from $\lambda$ 3700 \AA\ to $\lambda$ 9200 \AA . Moderate to high spectral resolutions have been required for the sample selection. Also they have been required to provide measurements of the auroral and nebular [SIII] emission lines at $\lambda$ 6312 \AA\ and $\lambda$ 9069 \AA\ respectively, necessary to derive [SIII] line electron temperatures and hence provide direct determinations of sulphur abundances. Most of the spectroscopy has been obtained with large telescopes providing very good quality data. Details can be found in the references listed in Tables \ref{RefDHR} and \ref{RefHII} that give the ID number of the object in column 1, some information about the instrumentation used in column 2 and the reference to the data source in column 3. 

The spectral information from extragalactic H II regions has been traditionally obtained from single-aperture or long-slit observations of the brightest part of the region, under the tacit assumption that the observations and the derived measurements are representative of the whole H II region, i.e. basically internal variations within the nebula are assumed to be minimal or non-existent.
\citet{2013MNRAS.430..472L} used IFS data for the centre and an external region of  M33  to test whether the description of the whole H II region with a single value for different parameters is valid and studied internal gradients of the intrinsic parameters of the H II regions for apertures or shells defined according to their average surface brightness in H$\alpha$. Cumulative fluxes of the [SII] and [SIII] lines for concentric individual shells turned out to be remarkly constant through both the high metallicity observed region and the low metallicity one leading to constant values of S$^+$/H$^+$, S$^{++}$/H$^+$ and S/H.

The reddening corrected intensities of the emission lines of interest for this work, relative to that of H$\beta$, together with their uncertainties, are listed in Tables \ref{linesDHR} and \ref{linesHII}.  

\begin{table*}
\centering
    \caption{References to published data sets for objects in the DHR sub-sample of the present compilation. This is a sample table consisting of the first 7 rows of data. The full table is available online.}
 \label{RefDHR}
\begin{tabular}{ccc}
\hline
ID  & Telescope+Instrument & Reference\\ \hline
1-2 & INT(2.5m)+IDS  & \citet{1987MNRAS.226...19D}\\ 
3 & INT(2.5m)+IDS  & \citet{1988MNRAS.235..633V}\\
4-7 & INT(2.5m)+IDS & \citet{1993MNRAS.260..177P}\\ 
8-12 & WHT(4.2m)+ISIS & \citet{1994ApJ...437..239G}\\ 
13 & MMT(4.5m, 6.5m)+ Blue Channel and Red Chanell Spectrograp & \citet{1994ApJ...426..123G}\\ 
14-16 & WHT(4.2m)+ISIS & \citet{1995ApJ...439..604G}\\ 
17-24 & INT(2.5m)+IDS and KPNO(2.1m)+ GoldCam CCD & \citet{1997ApJ...489...63G}\\ 
\hline     
\end{tabular}
\end{table*}
\begin{table*}
\centering
    \caption{References to published data sets for objects in the \HII Gal sub-sample of the present compilation. This is a sample table consisting of the first 7 rows of data. The full table is available online.}
 \label{RefHII}
\begin{tabular}{cccc}
\hline
ID & Telescope+Instrument & Reference\\ \hline

257-258 & Large Cass  McDonal(2.7m)+MMT(6.2m) Spectrograph & \citet{1993ApJ...411..655S}\\ 
259 & ISIS+WHT (4.2m) + MMT(6.2m) & \citet{1994ApJ...431..172S}\\ 
260-270 & IDS+INT(2.5m) & \citet{2003MNRAS.346..105P}\\ 
271-273 & ISIS+WHT(4.2m) & \citet{2006MNRAS.372..293H}\\ 
274-277 & UVES+VLT(8.2m) & \citet{2007ApJ...656..168L}\\ 
278-284 & ISIS+WHT(4.2m) & \citet{2008MNRAS.383..209H}\\ 
285-287 & FORS+UVES+VLT(8.2m) & \citet{2009AandA...503...61I}\\ 
\hline     
\end{tabular}
\end{table*}
\begin{table*}
\centering
\caption{Reddening corrected emission line intensities, relative to H$\beta$ = 100 for objects in the DHR sub-sample. This is a sample table consisting of the first 7 rows of data. The full table is available online. }
\label{linesDHR}
\begin{tabular}{cccccccc}
\cline{1-8}
  & Line & [OII] & [OIII] & [OIII] & [SIII] & [SII] & [SIII]\\
  & $\lambda$ & 3727, 3729 & 4363 &4959, 5007 & 6312  & 6717, 6731 & 9069, 9532\\\hline
\multicolumn{1}{c}{ID} & \multicolumn{1}{c}{Region} & \multicolumn{6}{c}{I($\lambda $)$^{a}$} \\ \hline
1 & N604D (M33) & 161.4 $\pm$ 2.0 & -& 340.4 $\pm$ 10.0 & 0.9 $\pm$ 0.4 & 23.4 $\pm$ 1.5 & 123.5 $\pm$ 2.5\\ 
2 & N604E & 257.0 $\pm$ 2.0 & -& 204.9 $\pm$ 3.3 & 1.1 $\pm$ 0.4 & 51.3 $\pm$ 2.9 & 127.5 $\pm$ 6.9\\
3 & NGC595 (M33) & 213.8 $\pm$ 10.01 & -& 189.6 $\pm$ 6.0 & 0.8 $\pm$ 0.15 & 24.9 $\pm$ 1.0 & 97.2 $\pm$ 1.6\\ 
4 & NGC3310-A & 259.0 $\pm$ 4.0 & 1.8 $\pm$ 0.5 & 305.0 $\pm$ 1.6 & 1.1 $\pm$ 0.3 & 49.0 $\pm$ 1.2 & 43.0 $\pm$ 0.7\\ 
5 & NGC3310-C & 316.0 $\pm$ 10.0 & 2.5 $\pm$ 0.7 & 326.0 $\pm$ 2.4 & 1.4 $\pm$ 0.2 & 55.0 $\pm$ 0.4 & 81.0 $\pm$ 0.9\\ 
6 & NGC3310-E & 313.0 $\pm$ 4.5 & -& 271.0 $\pm$ 1.5 & 1.4 $\pm$ 0.7 & 62.0 $\pm$ 2.7 & 76.0 $\pm$ 2.4\\ 
7 & NGC3310-L & 469.0 $\pm$ 19.0 & -& 225.0 $\pm$ 8.0 & 1.1 $\pm$ 0.5 & 85.0 $\pm$ 2.0 & 67.0 $\pm$ 3.5\\ 
\hline
\end{tabular}
\begin{tablenotes}
\centering
\item $^a$ Values normalized to I(H$\beta$)/100. 
\end{tablenotes}
\end{table*}

\begin{table*}
\centering
\caption{Reddening corrected emission line intensities, relative to H$\beta$ = 100, for objects in the \HII Gal sub-sample. This is a sample table consisting of the first 7 rows of data. The full table is available online.}
\label{linesHII}
\begin{tabular}{cccccccccc}
 \cline{1-10}
  & Line & [OII] & [OIII] & [ArIV] & [OIII] & [SIII] & [SII] & [ArIII] & [SIII]\\
  & $\lambda$ & 3727, 3729 & 4363 & 4740 & 4959, 5007 & 6312  &  6717, 6731 & 7136 & 9069, 9532\\\hline
\multicolumn{1}{c}{ID} & \multicolumn{1}{c}{Region} & \multicolumn{8}{c}{I($\lambda $)$^{a}$} \\ \hline
257 & IZW18NW & 26.4 $\pm$ 1.1 & 4.4 $\pm$ 0.7 & -& 259.8 $\pm$ 7.3 & 1.27 $\pm$ 0.03 & 3.58 $\pm$ 0.29 & -& 10.32 $\pm$ 0.28\\ 
258 & IZW18SE & 46.6 $\pm$ 2.2 & 6.2 $\pm$ 0.4 & -& 232.1 $\pm$ 6.7 & 0.63 $\pm$ 0.05 & 7.05 $\pm$ 0.11 & -& 13.42 $\pm$ 0.34\\ 
259 & UGC4483 & 83.5 $\pm$ 5.7 & 6.6 $\pm$ 0.4 & -& 362.8 $\pm$ 9.4 & 0.57 $\pm$ 0.05 & 12.43 $\pm$ 0.16 & -& 28.6 $\pm$ 4.3\\ 
260 & Mrk5 & 212.9 $\pm$ 3.1 & 4.4 $\pm$ 0.5 & -& 511.3 $\pm$ 6.3 & 1.39 $\pm$ 0.04 & 37.8 $\pm$ 0.45 & -& 48.1 $\pm$ 3.4\\ 
261 & 0749+568 & 166.8 $\pm$ 4.3 & 9.8 $\pm$ 1.1 & -& 655.2 $\pm$ 13.8 & 1.6 $\pm$ 0.2 & 23.4 $\pm$ 0.2 & -& 43.688 $\pm$ 4.816\\ 
262 & 0926+606 & 178.5 $\pm$ 1.2 & 8.3 $\pm$ 0.3 & -& 640.0 $\pm$ 3.6 & 2.3 $\pm$ 0.3 & 32.0 $\pm$ 0.3 & -& 48.2 $\pm$ 5.8\\ 
263 & Mrk709 & 181.6 $\pm$ 3.5 & 8.8 $\pm$ 0.5 & -& 522.6 $\pm$ 7.0 & 2.0 $\pm$ 0.1 & 62.3 $\pm$ 0.5 & -& 29.928 $\pm$ 3.44\\ 
\hline                                              
\end{tabular}
\begin{tablenotes}
\centering
\item $^a$ Values normalized to I(H$\beta$)/100. 
\end{tablenotes}
\end{table*}

\section{Direct derivation of abundances}

The physical conditions -- electron temperatures, electron density and abundances -- for the whole sample have been recalculated using PyNeb \citep{2015A&A...573A..42L}, based on a five-level statistical equilibrium model, using the atomic coefficients given in Table 3 of \citet{2018MNRAS.478.5301F}. According to the different data sources used, electron densities were determined from the [SII] $\lambda$ 6717{\AA} / $\lambda$ 6731{\AA} line ratio yielding small values, well below critical, in most cases, therefore a value of $n_e = 100$ $cm^{-3}$ has been assumed through this work.

 \begin{figure}
\centering
 \includegraphics[width=\columnwidth]{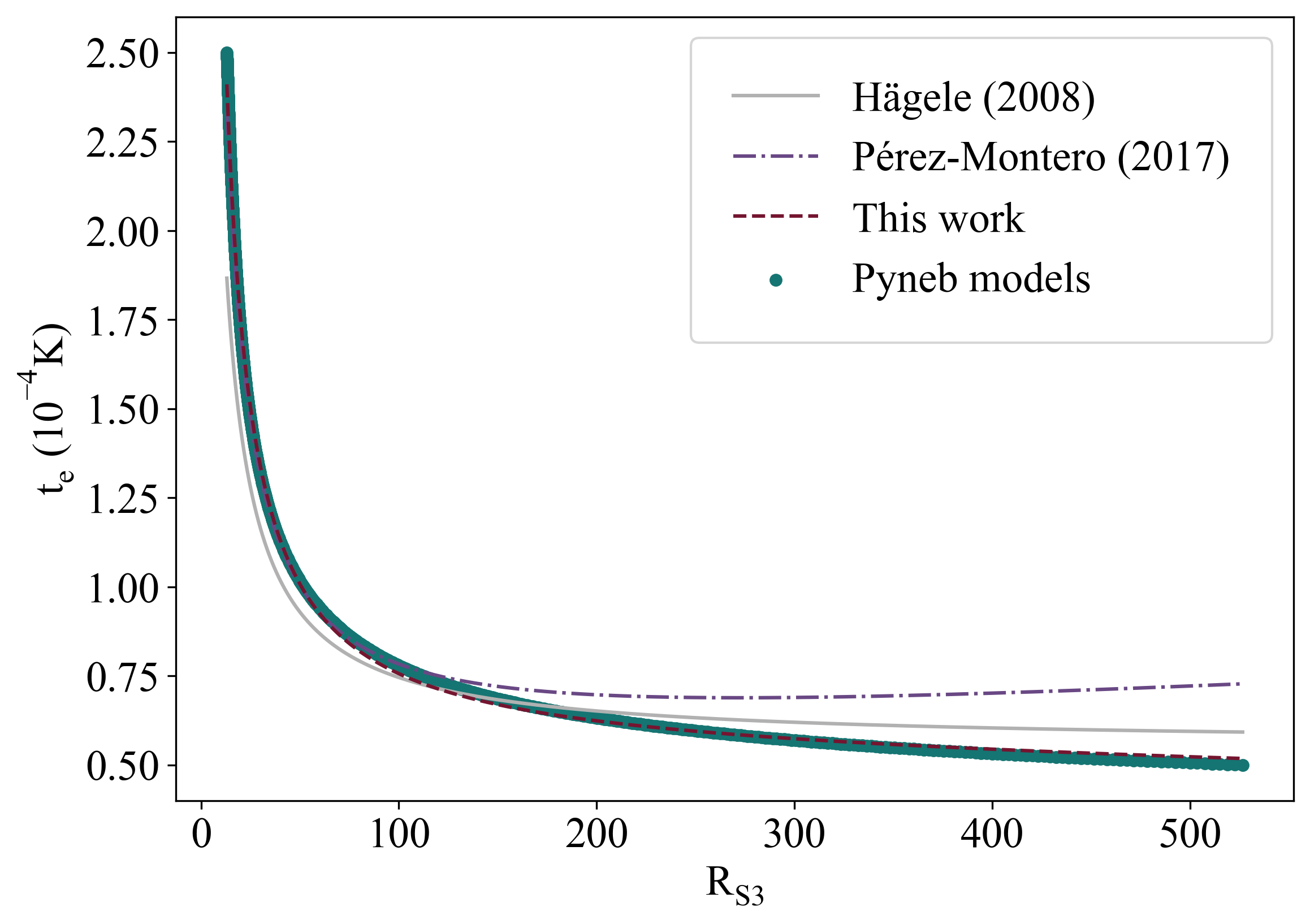}
 \caption{The sulphur temperature, T$_e$([SIII]) as a function of the line ratio R$_{S3}$, defined in the text, derived using the PyNeb package.}
 \label{teS3_Pyneb}
\end{figure} 

\begin{figure}
\centering
 \includegraphics[width=\columnwidth]{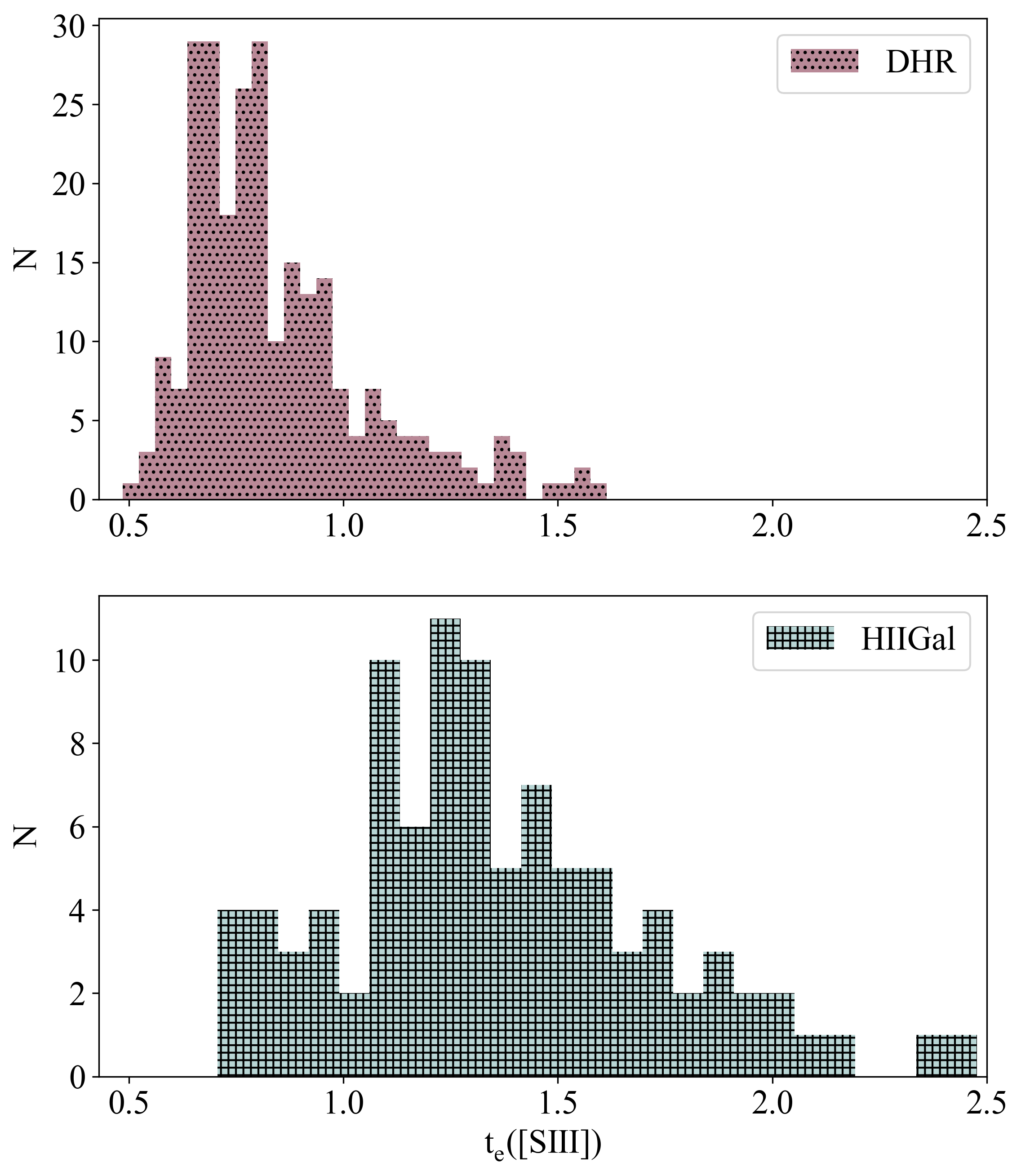}
 \caption{Distribution of the derived sulphur temperature, t$_e$([SIII]), for the two sub-samples studied, as labelled.}
 \label{hist_te}
\end{figure}

The S$^{++}$ electron temperature, $T_e([SIII])$ has been derived from the ratio of intensities of the nebular to auroral [SIII] lines ($R_{S3}= I(\lambda$ 9069{\AA}+$\lambda$ 9532{\AA})/I($\lambda$ 6312{\AA})). Since most expressions quoted in the literature do not apply to electron temperatures lower than about 8000 K, too high for many disc \HII\ regions present in our sample, we have used PyNeb to calculate $t_e([SIII])$ in the required temperature extended range. The results can be seen in Figure \ref{teS3_Pyneb} where it is shown how the expressions commonly used, for example the ones given in \citet{2008MNRAS.383..209H} or \citet{2017PASP..129d3001P}, overestimate the electron temperature in these cases thus leading to lower derived sulphur abundances. The fit to the PyNeb calculations, shown in Figure \ref{teS3_Pyneb}, give:

\begin{equation}
t_{e}([SIII]) = 0.5597-1.808\cdot 10^{-4} R_{S3}+\frac{22.66}{R_{S3}} 
\label{eq_1}
\end{equation}

\noindent where $t_{e}([SIII]) = 10^{-4} \times T_e([SIII])$. This expression is valid for 5000 K $\leq$ T$_e$([SIII]) $\leq$ 25000 K and shows a very weak dependence on electron density, increasing by about 3\% when the latter increases from 100 to 1000 cm${-3}$ \citep{2017PASP..129d3001P}. The derived t$_e$([SIII]) values together with their computed errors are listed in column 2 of Tables \ref{ResHDR} and \ref{SresHII} for each object of sub-samples DHR and \HII{Gal} respectively whose ID numbers are listed in column 1, and their corresponding distributions can be seen in Figure \ref{hist_te}.

\begin{table*}
\centering
\caption{Sulphur electron temperature and ionic abundances of objects in sub-sample DHR. This is a sample table consisting of the first 7 rows of data. The full table is available online.}
\label{ResHDR}
\begin{tabular}{ccccc}
\hline
ID & t$_e$([SIII])$^a$ & 12+log(S$^{+}$/H$^{+}$)&	12+log(S$^{++}$/H$^{+}$) & 12+log((S$^{+}$+S$^{++}$)/H$^{+}$)
\\ \hline
1 & 0.700 $\pm$ 0.074 & 6.22 $\pm$ 0.14 & 7.13 $\pm$ 0.1 & 7.18 $\pm$ 0.08\\ 
2 & 0.734 $\pm$ 0.072 & 6.49 $\pm$ 0.12 & 7.09 $\pm$ 0.09 & 7.19 $\pm$ 0.07\\ 
3 & 0.724 $\pm$ 0.035 & 6.2 $\pm$ 0.06 & 6.99 $\pm$ 0.04 & 7.05 $\pm$ 0.04\\ 
4 & 1.132 $\pm$ 0.158 & 5.96 $\pm$ 0.11 & 6.19 $\pm$ 0.09 & 6.39 $\pm$ 0.06\\ 
5 & 0.941 $\pm$ 0.056 & 6.21 $\pm$ 0.06 & 6.63 $\pm$ 0.04 & 6.77 $\pm$ 0.03\\ 
6 & 0.967 $\pm$ 0.209 & 6.23 $\pm$ 0.2 & 6.58 $\pm$ 0.15 & 6.74 $\pm$ 0.11\\ 
7 & 0.921 $\pm$ 0.17 & 6.42 $\pm$ 0.18 & 6.57 $\pm$ 0.13 & 6.8 $\pm$ 0.1\\ 
\hline
\end{tabular}
\begin{tablenotes}
\centering
\item $^a$ In units of 10$^{4}$K.\\
\end{tablenotes}
\end{table*}

\begin{table*}
\centering
\caption{Sulphur electron temperature and ionic abundances of objects in sub-sample \HII Gal. This is a sample table consisting of the first 7 rows of data. The full table is available online.}
\label{SresHII}
\begin{tabular}{ccccc}
\hline
ID  & t$_e$([SIII])$^a$ & 12+log(S$^{+}$/H$^{+}$)&	12+log(S$^{++}$/H$^{+}$) &  12+log((S$^{+}$+S$^{++}$)/H$^{+}$)
\\ \hline
257 & 1.94 $\pm$ 0.116 & 4.39 $\pm$ 0.03 & 5.18 $\pm$ 0.03 & 5.25 $\pm$ 0.02\\ 
258 & 1.518 $\pm$ 0.088 & 4.86 $\pm$ 0.04 & 5.46 $\pm$ 0.03 & 5.56 $\pm$ 0.03\\ 
259 & 1.562 $\pm$ 0.154 & 5.08 $\pm$ 0.06 & 5.77 $\pm$ 0.08 & 5.85 $\pm$ 0.06\\ 
260 & 1.308 $\pm$ 0.109 & 5.71 $\pm$ 0.06 & 6.12 $\pm$ 0.05 & 6.27 $\pm$ 0.04\\ 
261 & 1.749 $\pm$ 0.204 & 5.27 $\pm$ 0.06 & 5.88 $\pm$ 0.07 & 5.97 $\pm$ 0.05\\ 
262 & 1.495 $\pm$ 0.123 & 5.53 $\pm$ 0.05 & 6.03 $\pm$ 0.06 & 6.15 $\pm$ 0.05\\ 
263 & 1.616 $\pm$ 0.144 & 5.76 $\pm$ 0.05 & 5.76 $\pm$ 0.06 & 6.06 $\pm$ 0.04\\ 
\hline
\end{tabular}
\begin{tablenotes}
\centering
\item $^a$ In units of 10$^{4}$K.\\
\end{tablenotes}
\end{table*}

Just for consistency, we have derived the O$^{++}$ temperature, T$_e$([OIII]), using also PyNeb, from the ratio of intensities of the nebular to auroral [OIII] lines ($R_{O3}$= I($\lambda$ 4959{\AA}+$\lambda$ 5007{\AA})/I($\lambda$ 4363{\AA})), when data on the former exist, as:

\begin{equation}
t_e([OIII]) = 0.8254 - 0.0002415 \cdot R_{O3} + \frac{47.77}{R_{O3}} 
\label{eq_2}
\end{equation}

\noindent where $t_{e}([OIII]) = 10^{-4} \times T_e([OIII])$. This expression is valid for 7000 K $\leq$ T$_e$([OIII]) $\leq$ 25000 K. The derived t$_e$([OIII]) values together with their computed errors are listed in column 2 of Table \ref{OresDHR} for the objects of sub-sample DHR for which the [OIII] $\lambda$4363 \AA\ has been measured, and Table \ref{OresHII} for each object of sub-sample HIIGal. In both tables the object ID number is given in column 1.

\begin{table*}
\centering
\caption{Oxygen electron temperature and ionic abundances of objects in sub-sample DHR. Only objects with measured [ OIII] $\lambda$ 4363 \AA\ line intensities are listed. This is a sample table consisting of the first 7 rows of data. The full table is available online.}
\label{OresDHR}
\begin{tabular}{ccccc}
\hline
ID & t$_e$([OIII])$^a$ & 12+log(O$^{+}$/H$^{+}$)&	12+log(O$^{++}$/H$^{+}$) & 12+log(O/H)
\\ \hline
4 & 1.066 $\pm$ 0.079 & 7.72 $\pm$ 0.21 & 7.34 $\pm$ 0.09 & 7.87 $\pm$ 0.13\\ 
5 & 1.16 $\pm$ 0.103 & 8.15 $\pm$ 0.11 & 7.23 $\pm$ 0.09 & 8.19 $\pm$ 0.09\\ 
8 & 1.604 $\pm$ 0.02 & 6.75 $\pm$ 0.08 & 6.85 $\pm$ 0.01 & 7.1 $\pm$ 0.03\\ 
9 & 1.55 $\pm$ 0.044 & 7.05 $\pm$ 0.28 & 6.9 $\pm$ 0.03 & 7.28 $\pm$ 0.14\\ 
10 & 1.52 $\pm$ 0.013 & 6.77 $\pm$ 0.05 & 6.9 $\pm$ 0.01 & 7.14 $\pm$ 0.02\\ 
11 & 1.495 $\pm$ 0.028 & 6.63 $\pm$ 0.15 & 6.92 $\pm$ 0.02 & 7.1 $\pm$ 0.05\\ 
12 & 1.536 $\pm$ 0.111 & 6.52 $\pm$ 0.25 & 6.91 $\pm$ 0.06 & 7.06 $\pm$ 0.08\\ 
\hline
\end{tabular}
\begin{tablenotes}
\centering
\item $^a$ In units of 10$^{4}$K.\\
\end{tablenotes}
\end{table*}

\begin{table*}
\centering
\caption{Oxygen electron temperature and ionic abundances of objects in sub-sample \HII Gal. This is a sample table consisting of the first 7 rows of data. The full table is available online.}
\label{OresHII}
\begin{tabular}{ccccc}
\hline
ID & t$_e$([OIII])$^a$ & 12+log(O$^{+}$/H$^{+}$)&	12+log(O$^{++}$/H$^{+}$) & 12+log(O/H)
\\ \hline
257 & 1.62 $\pm$ 0.133 & 6.0 $\pm$ 0.06 & 7.25 $\pm$ 0.07 & 7.27 $\pm$ 0.06\\ 
258 & 2.092 $\pm$ 0.091 & 6.54 $\pm$ 0.07 & 6.95 $\pm$ 0.03 & 7.09 $\pm$ 0.03\\ 
259 & 1.681 $\pm$ 0.058 & 6.75 $\pm$ 0.11 & 7.35 $\pm$ 0.03 & 7.45 $\pm$ 0.03\\ 
260 & 1.208 $\pm$ 0.05 & 7.41 $\pm$ 0.11 & 7.88 $\pm$ 0.04 & 8.01 $\pm$ 0.04\\ 
261 & 1.524 $\pm$ 0.083 & 6.92 $\pm$ 0.11 & 7.71 $\pm$ 0.05 & 7.78 $\pm$ 0.04\\ 
262 & 1.426 $\pm$ 0.023 & 7.14 $\pm$ 0.09 & 7.78 $\pm$ 0.01 & 7.87 $\pm$ 0.02\\ 
263 & 1.615 $\pm$ 0.048 & 7.05 $\pm$ 0.09 & 7.55 $\pm$ 0.02 & 7.67 $\pm$ 0.03\\ 
\hline
\end{tabular}
\begin{tablenotes}
\centering
\item $^a$ In units of 10$^{4}$K.\\
\end{tablenotes}
\end{table*}

The atomic coefficients used for the derivation of electron temperatures have been taken from \citet{2018MNRAS.478.5301F}.

Since different ions are formed in regions of different temperatures, a certain ionisation structure for the nebula has to be assumed in each case, which differs for the objects of each sub-sample of this study, as described below.

\subsection{Sulphur abundances}

In the majority of objects of the DHR sub-sample most of the sulphur is expected to be in the form of S$^+$ and S$^{++}$, with S$^{++}$ being the dominant ionization stage, except perhaps in the highest metallicity cases. Due to the low ionisation potential for sulphur, the contribution by S$^0$ can be neglected, but, on the other hand, a certain contribution by S$^{3+}$ is expected in high excitation, low metallicity, objects which amount only to about 4\% of the sub-sample.

\begin{figure}
\centering
 \includegraphics[width=\columnwidth]{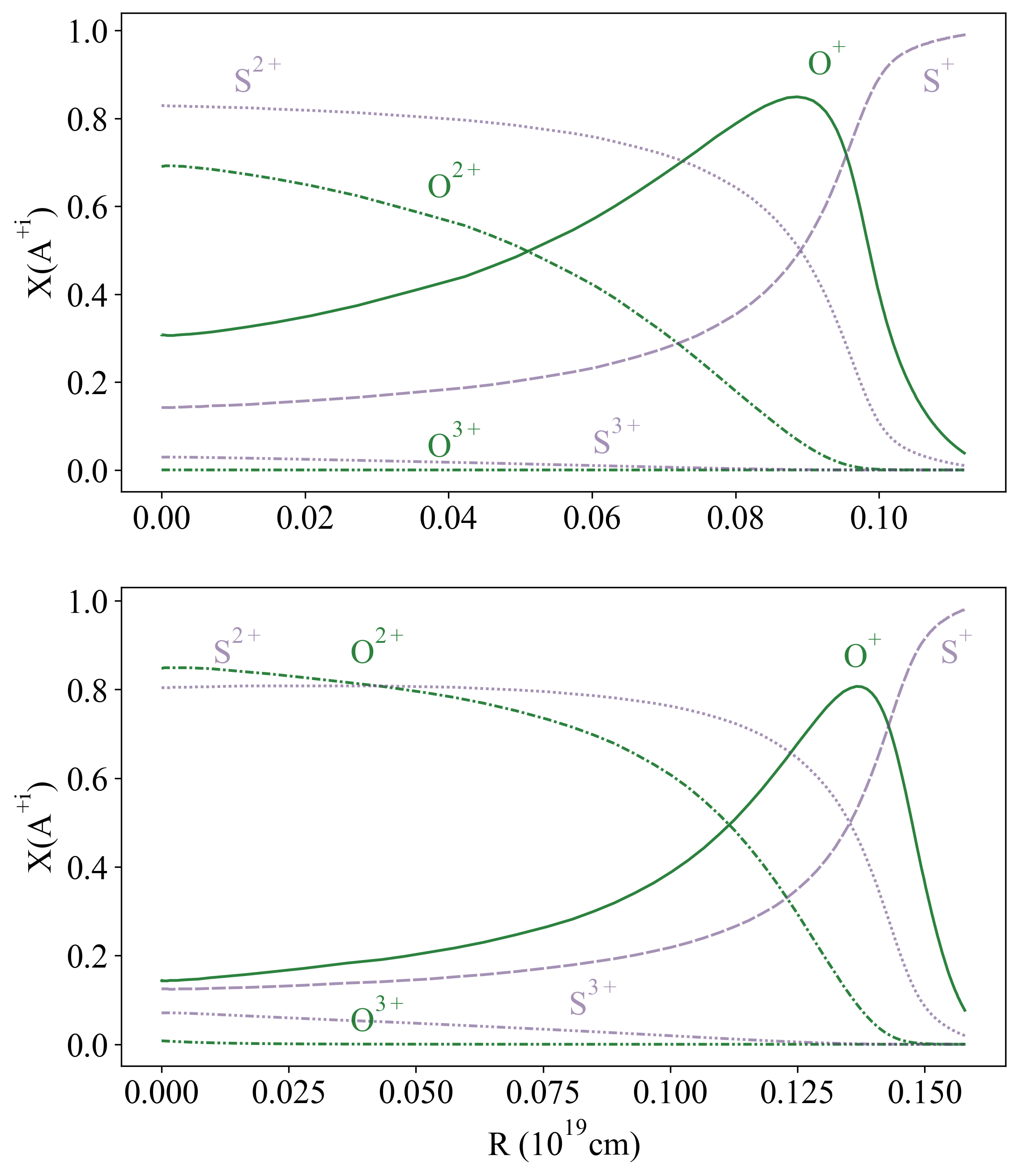}
 \caption{Upper panel: ionisation structure of a Cloudy model \citep[version 17.01][]{2017RMxAA..53..385F} of low density (100 cm$^{-3}$) HII regions of half solar abundance ionised by a 5 Ma star cluster, representative of the objects in the DHR sub-sample. Lower panel: ionisation structure of \HII\ regions of low metallicity gas  (Z=0.004) ionised by a young (1 Ma) star cluster of 10$^5$ M$_{\odot}$, representative of the objects in sub-sample \HII\ Gal .}
 \label{ionstruc}
\end{figure}

Figure \ref{ionstruc} shows the ionisation structure computed from a Cloudy model \citep[version 17.01][]{2017RMxAA..53..385F} of low density (100 cm$^{-3}$) \HII\ regions with half solar abundance ionised by a 5 Ma star cluster (upper panel) and a \HII\ galaxy of low metallicity  (Z=0.004) ionised by a young (1 Ma) star cluster of 10$^5$ M$_{\odot}$ (lower panel), showing the contribution by S$^{3+}$. The star cluster spectral energy distributions have been synthesised with the PopStar code \citep{2009MNRAS.398..451M}.  In the two cases, both S$^+$ and S$^{++}$ ions can be seen to to overlap in a transition zone that encompasses almost the whole nebula, a result that has also been shown, for example, by \citet{1992AJ....103.1330G} and \citet{2012ApJ...755...40P}.

According to this, we have assumed an only zone in which $T_e(S^+) \approx T_e(S^{++}) = T_e([SIII])$. $S^+$ and $S^{++}$ ionic abundances have been calculated using the expressions derived using the PyNeb package \citep{2015A&A...573A..42L} and the atomic coefficients listed in Table 2 of \citet{2018MNRAS.478.5301F}:

\begin{equation}
\begin{split}
12+log\left(\frac{S^{+}}{H^{+}}\right)=log\left(\frac{I(\lambda 6717\AA +\lambda 6731\AA)}{I(H_{\beta})}\right)+5.516+ \\
+\frac{0.884}{t_{e}([SIII]) }-0.480log(t_{e}([SIII]) 
\end{split}
\label{eq_3}
\end{equation}

\begin{equation}
\begin{split}
12+log\left(\frac{S^{2+}}{H^{+}}\right)=log\left(\frac{I(\lambda 9069\AA+\lambda 9532\AA)}{I(H_{\beta})}\right)+6.059+\\
+\frac{0.608}{t_{e}([SIII]) }-0.706log(t_{e}([SIII]) 
\end{split}
\label{eq_4}
\end{equation}

\noindent where I($\lambda $6717\AA\ + $\lambda $6731\AA) denotes the sum of the intensities of the two red [SII] lines, I($\lambda $9069\AA\ + $\lambda $9532\AA) denotes the sum of the two near infrared [SIII] lines \begin{footnote}{In the cases in which no data are given on the $\lambda$9532 \AA\ line a theoretical ratio I$_{9532}$([SIII])/I$_{9069}$([SIII])=2.44 has been assumed.}\end{footnote}, I(H$\beta$) denotes the H$\beta$ intensity and t$_e$([SIII]) denotes the [SIII] electron temperature in units of 10$^{-4}$ K. The derived ionic abundances, in the format $12+log(X^i/H^+)$, are given in columns 2 and 3 of Tables \ref{ResHDR} and \ref{SresHII} for each of the objects in the two studied sub-samples, whose ID numbers are listed in column 1. Column 4 of the tables gives the sum of the S$^+$ and S$^{++}$ relative to H$^+$. which, for the objects in the DHR sub-sample, can be taken to be the total S/H abundance, as we will see later.


On the other hand, the \HII\ Gal sub-sample includes a substantial proportion of low metallicity and high excitation objects for which the contribution by S$^{3+}$ has to be taken into account. A direct determination of this contribution would require the measurement of the flux from the infrared [SIV] line at 10.54 $\mu$m. In the absence of these data, as is the case for most of the objects of this sub-sample, the knowledge of ionisation correction factors (ICF) is required. The ICF for sulphur is defined as: 

\[ ICF(S^++S^{++}) = \frac{S}{H} \cdot \left(\frac{S^{+}+S^{++}}{H^+}\right)^{-1} \]

In order to calculate the individual ICF for our objects, we have used the relation between ionic ratios of Ar and S in contiguous ionisation stages: $Ar^{+2}/Ar^{+3}$ vs. $S^{2+}/S^{3+}$ . This relation has been calculated using the Cloudy photoionisation grid described in \citet{2018MNRAS.478.5301F} and is shown in their Figure 3. The linear fitting obtained is:

\begin{equation}
log\left(\frac{Ar^{2+}}{Ar^{3+}}\right) =(1.162\pm0.006) \cdot log\left(\frac{S^{2+}}{S^{3+}}\right)+(0.05\pm0.01) 
\label{eq_5}
\end{equation}

The application of this alternative requires the measurement of the [ArIII] and [ArIV] emission lines at  $\lambda\lambda$ 7135 and 4740 \AA\ respectively. Also, the [ArIV] lines, as well as those of [SIV], originate in a high excitation zone better characterised by the $O^{++}$ temperature (T$_e$([OIII]). 

The corresponding $\frac{Ar^{++}}{H^{+}}$ and $\frac{Ar^{3+}}{H^{+}}$ ionic ratios have been derived using the expressions: 

\begin{equation}
\begin{split}
12+log\left(\frac{Ar^{++}}{H^{+}}\right)=log\left(\frac{I(7135)}{I(H_{\beta})}\right)+6.145+ \\
+\frac{0.810}{t_{e}([SIII])}-0.502log(t_{e}([SIII])
\end{split}
\label{eq_6}
\end{equation}

\begin{equation}
\begin{split}
12+log\left(\frac{Ar^{3+}}{H^{+}}\right)=log\left(\frac{I(4740)}{I(H_{\beta})}\right)+6.362+\\
+\frac{1.174}{t_{e}([OIII])}-0.820log(t_{e}([OIII]) 
\end{split}
\label{eq_7}
\end{equation}
 
 \noindent using the references given above.
 
 The derived ICF for the objects in sub-sample \HII Gal for which measurements of the [ArIV] line exist are given in column 5 of Table \ref{ICF} together with the $Ar^{++}/H^+$, $Ar^{3+}/H^+$,and $S^{3+}/S^{++}$ ionic ratios, in columns 2,3 and 4 respectively. 
 
\begin{table*}
\centering
\caption{Ionisation correction factors and total S/H abundance for the objects in the general sample. Only objects with ICF > 1.1 are listed (see text for dateils). This is a sample table consisting of the first 7 rows of data. The full table is available online.}
\label{ICF}
\begin{tabular}{cccccccc}
\hline
ID & 12+log(Ar$^{++}$/H$^{+}$)&	12+log(Ar$^{3+}$/H$^{+}$) & log(S$^{3+}$/S$^{++}$) & ICF(Ar) & ICF($\eta$')& 12+log(S/H)
\\ \hline
4 & - & - & - & -& 1.17 $\pm$ 0.01 & 6.46 $\pm$ 0.07\\ 
8 & - & - & - & -& 1.82 $\pm$ 0.08 & 6.26 $\pm$ 0.05\\ 
9 & - & - & - & -& 1.38 $\pm$ 0.16 & 6.18 $\pm$ 0.15\\ 
10 & - & - & - & -& 1.48 $\pm$ 0.04 & 6.12 $\pm$ 0.04\\
11 & - & - & - & -& 1.64 $\pm$ 0.22 & 6.12 $\pm$ 0.14\\
12 & - & - & - & -& 1.71 $\pm$ 0.34 & 6.21 $\pm$ 0.21\\
13 & - & - & - & -& 1.16 $\pm$ 0.07 & 6.05 $\pm$ 0.08\\
 \hline
\end{tabular}
\end{table*}

 Since the ionisation of S$^{3+}$ requires high energy photons, it could be expected for the ICF to show a certain dependence on the hardness of the ionising radiation field. This can be mapped using the $\eta$ parameter defined by \citet{1988MNRAS.231..257V} as: 

\[ \eta = \frac{O^{+}/O^{++}}{S^{+}/S^{++}} \]

\noindent or its empirical analog: 

\[ \eta^{'} = \frac{[OII]/[OIII]}{[SII]/[SIII]} \]
 
\noindent where [OII], [OIII], [SII] and [SIII] stand for the sum of the two corresponding strong lines of each ion.

\begin{figure}
 \includegraphics[width=\columnwidth]{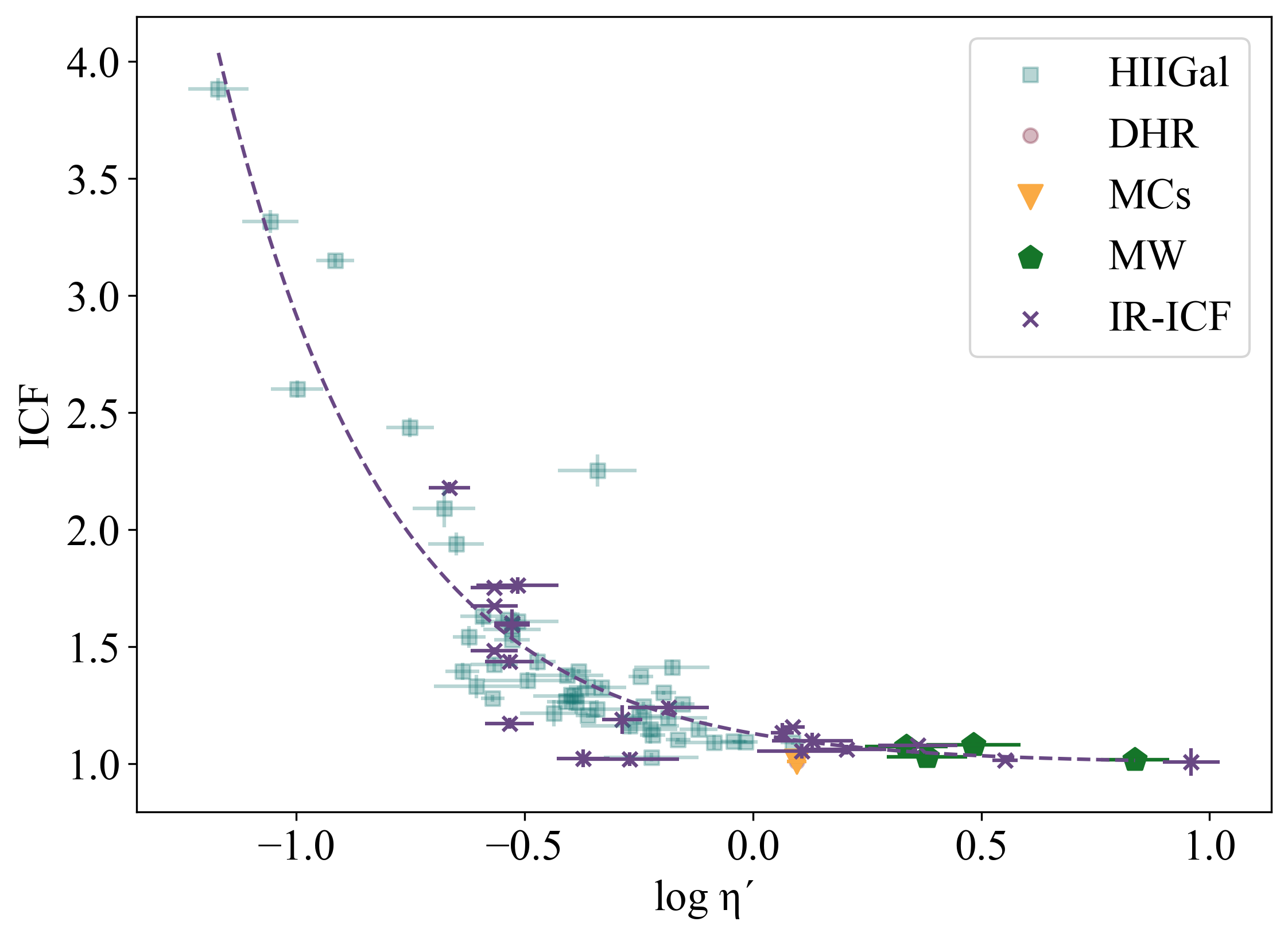}
 \caption{Relation between our derived ICF and the parameter $\eta$'. As can be seen, the ICF has a value $\simeq$ 1 for $\eta$'> 1.0.}
 \label{fig:ICF-eta}
\end{figure}

The relation between the derived ICF and $\eta$' is shown in Figure \ref{fig:ICF-eta} where it can be seen that the ICF has a value of 1.0, between the uncertainties, for values of $log\eta$' larger than 0.0 (T$_{ion} \approx$ 38000~ K) and increases as log$\eta$' decreases according to the expression: 
 
 \begin{equation}
log(ICF-1) = (-0.914\pm 0.064) + (-1.194 \pm 0.066)\cdot log \eta '
\label{eq_8}    
\end{equation}
 
 \noindent We have used this empirically derived function to estimate the ICF for the objects for which no data on the [ArIV] line exist and their values are given in column 6 of Table \ref{ICF} for the objects with ICF > 1.1. 
 
On the other hand, data obtained by the Spitzer satellite on the IR [SIV] and [SIII] emission lines, at 10.54 and 18.71 $\mu$m respectively, exist for some objects of our sample. Some very well studied objects, among which are included the two lowest metallicity observed galaxies: IZw18 and SBS0335-052E, have data from two different sources, which can serve to estimate the magnitude of the internal errors in this kind of observations. From these data we have derived the abundance ratio ${S^{3+}}/{S^{2+}}$ using the expression given in \cite{2016MNRAS.456.4407D}:

\begin{equation}
\begin{split}
log\left(\frac{S^{3+}}{S^{2+}}\right)=log\left(\frac{I(\lambda 10.54\mu m)}{I(\lambda 18.71\mu m)}\right)-0.833+ \\
+\frac{0.098}{t_{e}([OIII])}-0.525log\left(t_{e}([OIII])\right)+\\
+\frac{0.056}{t_{e}([SIII]}+0.639log\left(t_{e}([SIII])\right)
\end{split}
\label{eq_9}
\end{equation}

\begin{figure}
\centering
 \includegraphics[width=\columnwidth]{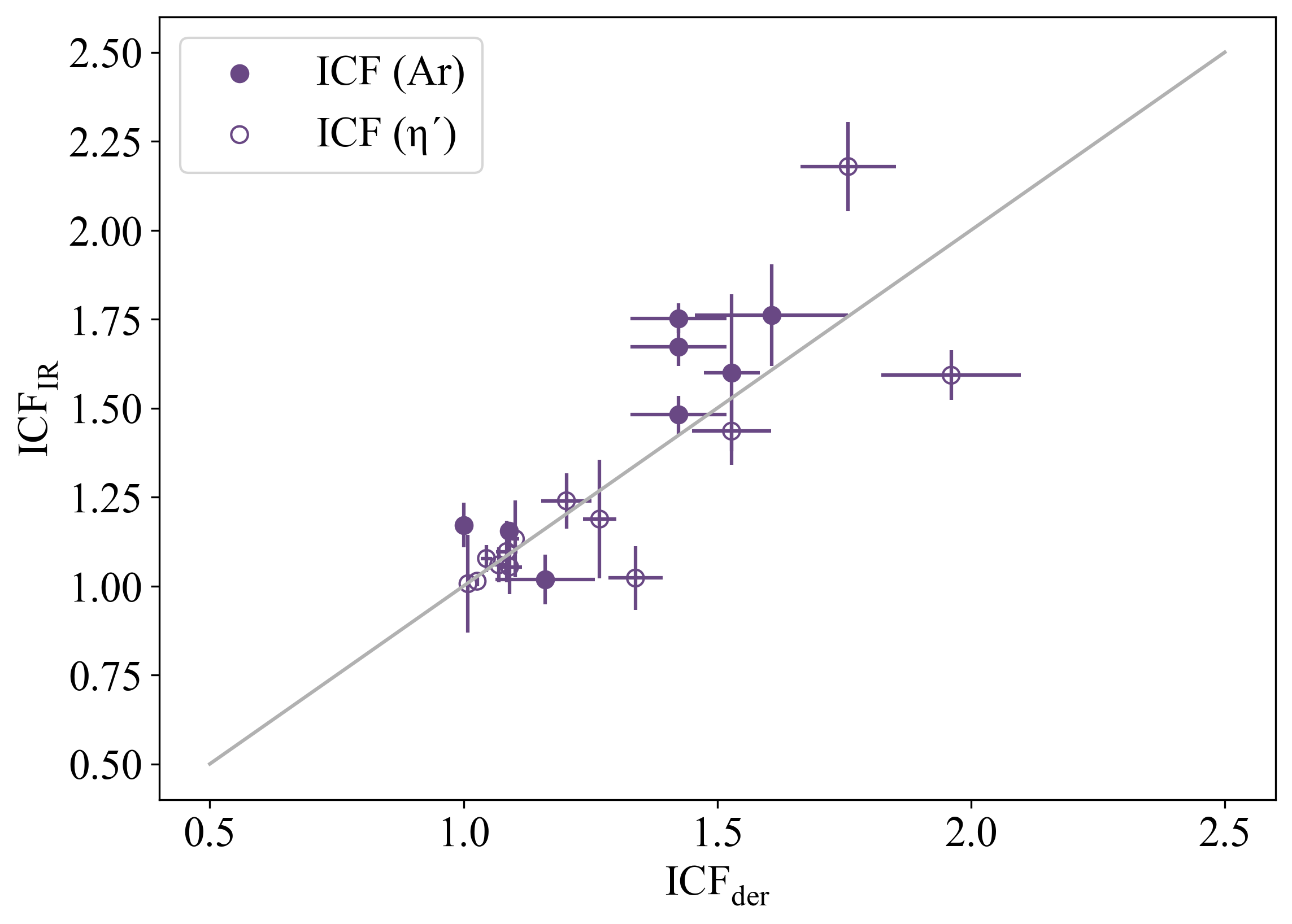}
 \caption{Comparison of the IR derived values of the sulphur ICF with the ones calculated in this work. For objects labelled as ICF(Ar) they have been calculated from the derived Ar$^{2+}$/Ar$^{3+}$ ratio and for the objects labelled ICF($\eta$') Eq. \ref{eq_8} has been used.}
 \label{fig:ICF}
\end{figure}

The infrared derived ICF values for all the objects in the general sample for which we have found suitable data (including some disc \HII\ regions in M101) are listed in Table \ref{ICF-IR} and Figure \ref{fig:ICF} shows a comparison of these IR derived values with the ones calculated using our method, either from the A$r^{+2}$/Ar$^{+3}$ ionic ratio or from Eq. \ref{eq_8} in the case in which no data in the [ArIV] lines exist. As can be seen, the agreement is rather good. We have therefore multiplied those values by the derived $(S^++S^{++})/H^+$ ratio to obtain the total S/H abundances which are given in column 7 of Table \ref{ICF}. 

\begin{table*}
\centering
\caption{IR derived ionisation correction factors (ICF). This is a sample table consisting of the first 7 rows of data. The full table is available online.}
\label{ICF-IR}
\begin{tabular}{cccccccc}
\hline
ID & [SIV]10.54$\mu$m & [SIII]18.71$\mu$m & log(S$^{++}$/S$^{3+}$) & ICF$_{IR}$ 
\\ \hline
32$^a$ & 28.39 $\pm$ 0.81 & 60.56 $\pm$ 3.94 & 1.05 $\pm$ 1.05 & 1.077 $\pm$ 0.038\\ 
35$^a$ & 2.75 $\pm$ 1.21 & 7.46 $\pm$ 1.49 & 1.17 $\pm$ 1.17 & 1.059 $\pm$ 0.049\\ 
38$^a$ & 3.33 $\pm$ 0.1 & 5.41 $\pm$ 0.64 & 0.91 $\pm$ 0.91 & 1.097 $\pm$ 0.087\\ 
40$^a$ & 3.08 $\pm$ 0.09 & 7.93 $\pm$ 1.68 & 1.17 $\pm$ 1.17 & 1.054 $\pm$ 0.076\\ 
45$^a$ & 12.92 $\pm$ 0.73 & 8.47 $\pm$ 1.38 & 0.56 $\pm$ 0.56 & 1.239 $\pm$ 0.078\\ 
173$^a$ & 0.18 $\pm$ 0.05 & 4.01 $\pm$ 0.18 & 2.0 $\pm$ 2.0 & 1.006 $\pm$ 0.137\\ 
257$^b$  & 0.48 $\pm$ 0.03 & 0.23 $\pm$ 0.02 & 0.3 $\pm$ 0.3 & 1.435 $\pm$ 0.095\\ 
\hline
\end{tabular}
\vspace{0.3cm}
\begin{tablenotes}
\centering
\item {\it Notes:} $^a$ \cite{2008ApJ...682..336G}; $^b$ \cite{2008ApJ...673..193W}; $^c$ \cite{2008ApJ...678..804E}; $^d$ \cite{2011ApJ...734...82I}.
\end{tablenotes}
\end{table*}

\subsection{Oxygen abundances}

In the case of oxygen, the most common procedure consists in assuming two ionisation zones: a high ionisation zone where O$^{++}$ originates and the electron temperature is well represented by T$_e$([OIII]), and a low ionization zone where O$^+$ originates and the electron temperature is well represented by T$_e$([OII]). T$_e$([OIII]) is derived from the ratio of intensities of the nebular to auroral [OIII] lines ($R_{O3}$= I($\lambda$4959{\AA} + $\lambda$5007{\AA}) / I($\lambda$4363{\AA})) as explained above, and T$_e$([OII]) is usually derived with the help of photoionisation models, \citep[e. g.][]{1990A&AS...83..501S,1992MNRAS.255..325P}:

\begin{equation}
T_e([OII])^{-1} = 0.50 \left(T_e([OIII])^{-1} + 0.00008\right) K  
\label{eq_10}
\end{equation}

Most of the oxygen is in the form of O$^+$ and O$^{2+}$, their relative contributions depending of the degree of ionisation. Only for exceptionally high excitation objects, those that show HeII emission at $\lambda$ 4686 \AA , there is a small contribution by O$^{3+}$ that in our case can be neglected. 

Then, the O$^+$/H$^+$ and  O$^{2+}$/H$^+$ abundance ratios are derived with the use of T$_e$([OII]) and  T$_e$([OIII]) respectively. A detailed description of this method, at times referred to as "the standard method", can be found in \citet{1994ApJ...435..647I}. However, T$_e$([OII]) depends on density so that the relation between  T$_e$([OII]) and T$_e$([OIII]) is not straightforward \citep[see Figure 8 of][]{2006MNRAS.372..293H} and it is clear that not a simple single relation exists between both. This is also the case when a homogeneous family of objects is considered, as can be seen for the sample of \HII\ galaxies analysed by \citet{2006A&A...448..955I}. Even though the graph shown in their Figure 4 is seen to be dominated by a  considerable scatter and  large error bars, most of the data with the smaller error bars fall off the relation between the temperatures assumed in the "standard method". Therefore, the determination of O$^+$/H$^+$ carries a large uncertainty. Unfortunately, O$^+$/H$^+$ can be of the same order as, or larger than, O$^{2+}$/H$^+$ in low excitation nebulae, and hence the resulting total value of the O/H abundance carries a large uncertainty for these objects. In our case this is circumvented since the assumption of $T_e([OII]) \simeq T_e([SII]) \simeq T_e([SIII])$ is made. The $O^+$ and $O^{++}$ ionic abundances have been calculated using the expressions: 

\begin{equation}
\begin{split}
12+log\left(\frac{O^{+}}{H^{+}}\right)=log\left(\frac{I(\lambda 3727+\lambda 3729)}{I(H_{\beta})}\right)+5.887+ \\
+\frac{1.641}{t_{e}([SIII])}-0.543log(t_{e}([SIII]) 
\end{split}
\label{eq_11}
\end{equation}

\begin{equation}
\begin{split}
12+log\left(\frac{O^{++}}{H^{+}}\right)=log\left(\frac{I(\lambda 4959+\lambda 5007)}{I(H_{\beta})}\right)+6.249+\\
+\frac{1.184}{t_{e}([OIII])}-0.708log(t_{e}([OIII]) 
\end{split}
\label{eq_12}
\end{equation}

\noindent and the references given above. The total O/H abundance is calculated as: 

\[\frac{O}{H}= \frac{O^{+}+O^{++}}{H^+}  \].

The derived ionic abundances for the two sub-samples used, in the format $12+log(X^i/H^+)$, are given in columns 2 and 3 of Tables \ref{OresDHR} and \ref{OresHII} for each of the objects in the two sub-samples for which data on the $\lambda$ 4363 \AA\ line are available. The ID object numbers are listed in column 1. Column 4 of the tables gives the total oxygen abundance: 12+log(O/H). Only objects with measurements of the $\lambda$ 4363 \AA\ line have been considered.


\section{Empirical Calibrations}

 
In the previous sections we have described the method to derive sulphur and oxygen abundances by the so called "direct method" that requires the actual determination of the gas electron temperature. However, in many cases, the emission lines necessary to do this are too faint to be detected, either due to the low brightness of the objects to observe or to the strong cooling effect of metals in some of the observed regions. In order to overcome this obstacle \citet{1979MNRAS.189...95P} and \citet{1979A&A....78..200A} pioneered methods to estimate the metallicity, usually traced by the oxygen abundance, using only strong, easily observable, emission lines.

The "strong line methods" assume that all \HII\ region spectra are essentially characterised by their overall metallicity, Z. However, other properties, mainly the hardness of the ionising radiation and the gas ionisation degree, are intimately related with it. The former can be inferred from the spectral energy distribution of the ionising source and can be parametrised by the $\eta$' parameter defined in Section 3.1 and the latter is represented by the ionisation parameter for hydrogen which is essentially the ratio of the ionising photon flux to the gas electron density and can be defined as: 

\[ u=\frac{Q(H)}{4\pi R^2\cdot c \cdot n_e} \]

\noindent where Q(H) is the flux of hydrogen ionising photons, R is the size of the ionised region and n$_e$ is the gas electron density (the speed of light,c,is included only to make the parameter dimensionless).  Since there will be a different ionisation parameter for each volume element at each point of the cloud, u is defined at the illuminated front of the cloud. This ionisation parameter can be adequately described by the emission line ratio $[SII]\lambda\lambda 6716,6731$ \AA /$[SIII]\lambda\lambda 9069,9532$ \AA\  \citep{1991MNRAS.253..245D}.
In fact, the three parameters: Z (as represented by the abundance of a chosen element as oxygen or sulphur, for example), $\eta$ (or $\eta^{'}$) and u are  non-linearly inter-linked but the exact nonlinear relation between them is still unknown. 

Observed line ratio diagnostics are central to the whole endeavour of metallicity determinations, hence the large amount of literature devoted to the identification of abundance indicators and their calibration \citep[see, for example][]{2005MNRAS.361.1063P,2013A&A...559A.114M}. In fact, there are two separate and important issues: the choice of the ideal indicator, and the proper abundance calibration. Although a plethora of different indicators have been proposed, it has not yet been possible to perform their calibration over the full range of metallicity with the necessary confidence due to a lack of data and an incomplete understanding of the physics in the high metallicity regime \citep[see][]{2004ApJ...615..228B,2006MNRAS.367.1139P}. The observation and measurement of the strong nebular [SIII] lines provides an alternative metallicity calibration, similar in nature to that of the commonly used R$_{23}$, but presenting the considerable advantage of remaining single-valued up to, at least, solar abundances.

\begin{figure}
 \includegraphics[width=\columnwidth]{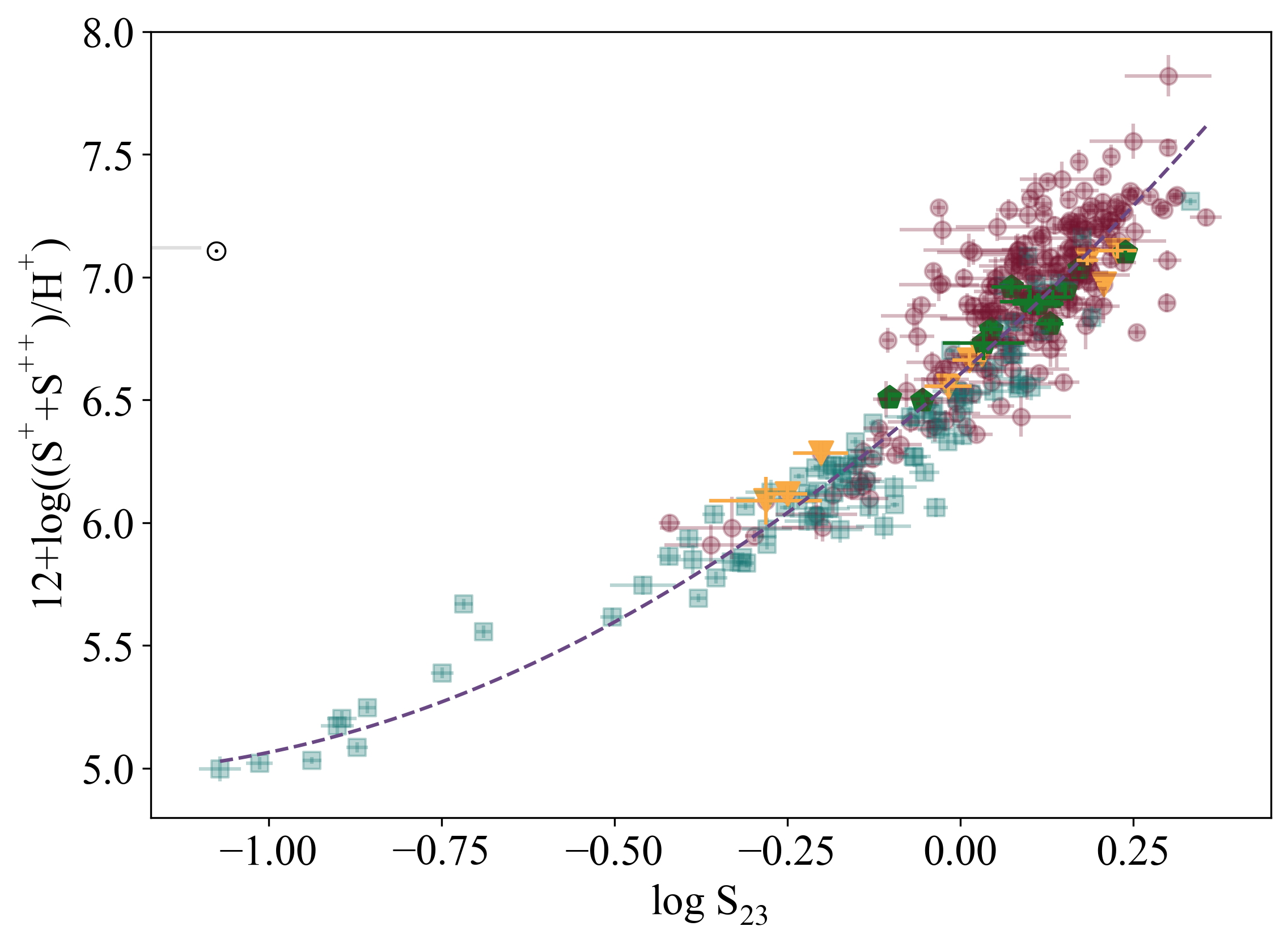}
 \includegraphics[width=\columnwidth]{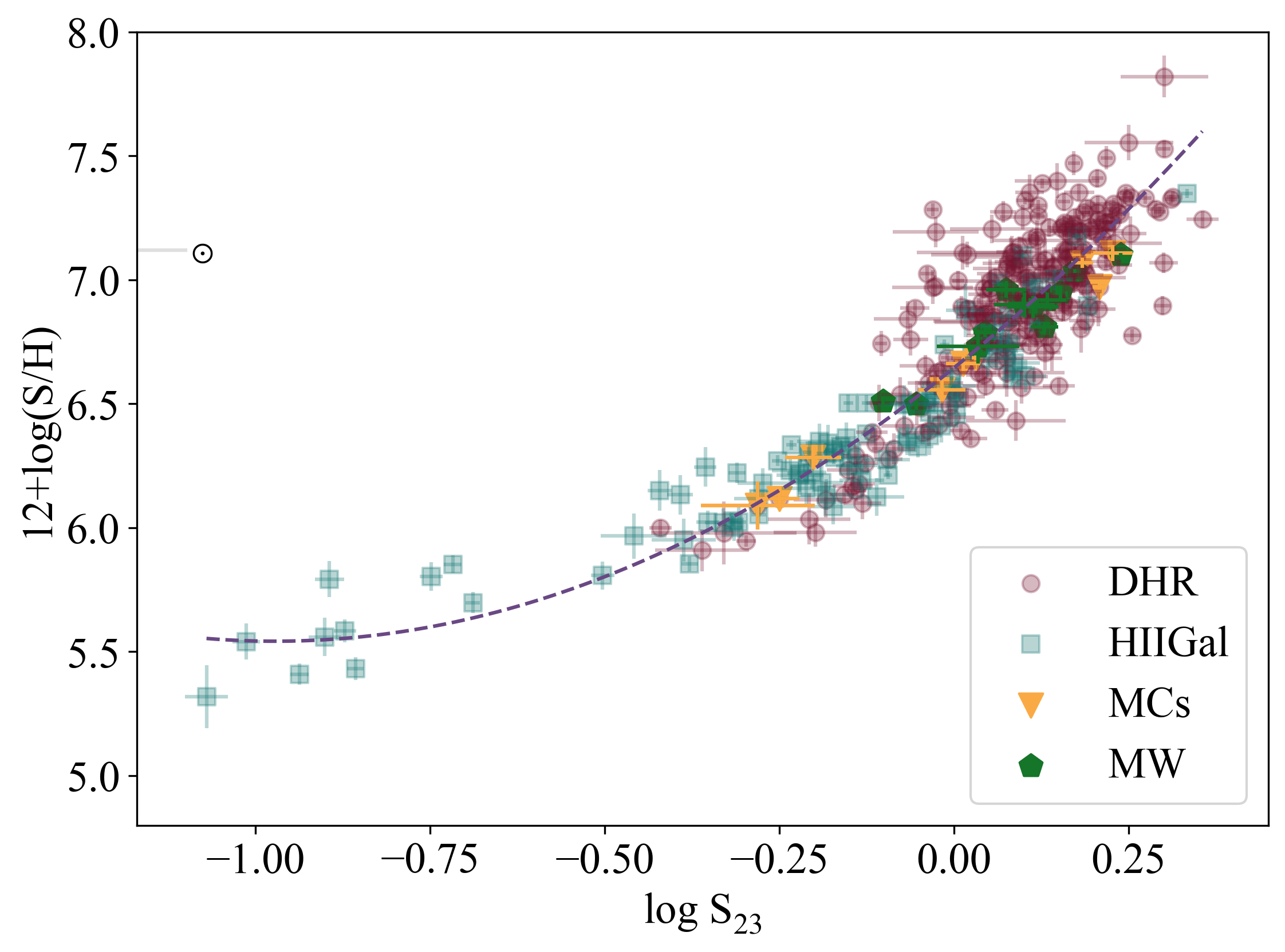}
 \caption{The calibration of the S$_{23}$ parameter in terms of the S$^+$+S$^{++}$/H$+$ ionic ratio (upper panel) and the total S/H abundance (lower panel).}
 \label{fig:calibracion}
\end{figure}

 The first comprehensive calibration of the S$_{23}$ parameter was presented in \citet{2000MNRAS.312..130D}. The sample of 196 objects was rather heterogeneous and only about 30\% of them had measurements of the weak auroral [SIII] line. During the last two decades, the number of objects with data on this line has increased along with the data quality, as corresponds to the use of more recent instrumentation and larger telescopes, which can improve the resulting calibration while reducing considerably the associated error, one of the objectives of the present work. The calibrations of the S$_{23}$ parameter in terms of the S$^+$+S$^{++}$/H$+$  ionic ratio (upper panel) and the total S/H abundance (lower panel) are shown in Figure \ref{fig:calibracion}. Second order polynomial fits yield the relations:

\begin{equation}
\begin{split}
12+log \left(\frac{S^+ + S^{++}}{H^+}\right)=(6.593\pm 0.010)+\\+ (2.439\pm 0.050)\cdot log S_{23} +(0.909\pm 0.076) \cdot (log S_{23})^2
\end{split}
\label{eq_13}    
\end{equation}

\begin{equation}
\begin{split}
12+log \left(\frac{S}{H}\right)=(6.636\pm 0.011)+\\+ (2.202\pm 0.050)\cdot log S_{23} +(1.060\pm 0.098) \cdot (log S_{23})^2
\end{split}
\label{eq_14}    
\end{equation}

\noindent with a typical deviation of 0.18 dex.

\section{Discussion}
\subsection{Characteristics of the observed objects}

\begin{figure}
\centering
 \includegraphics[width=\columnwidth]{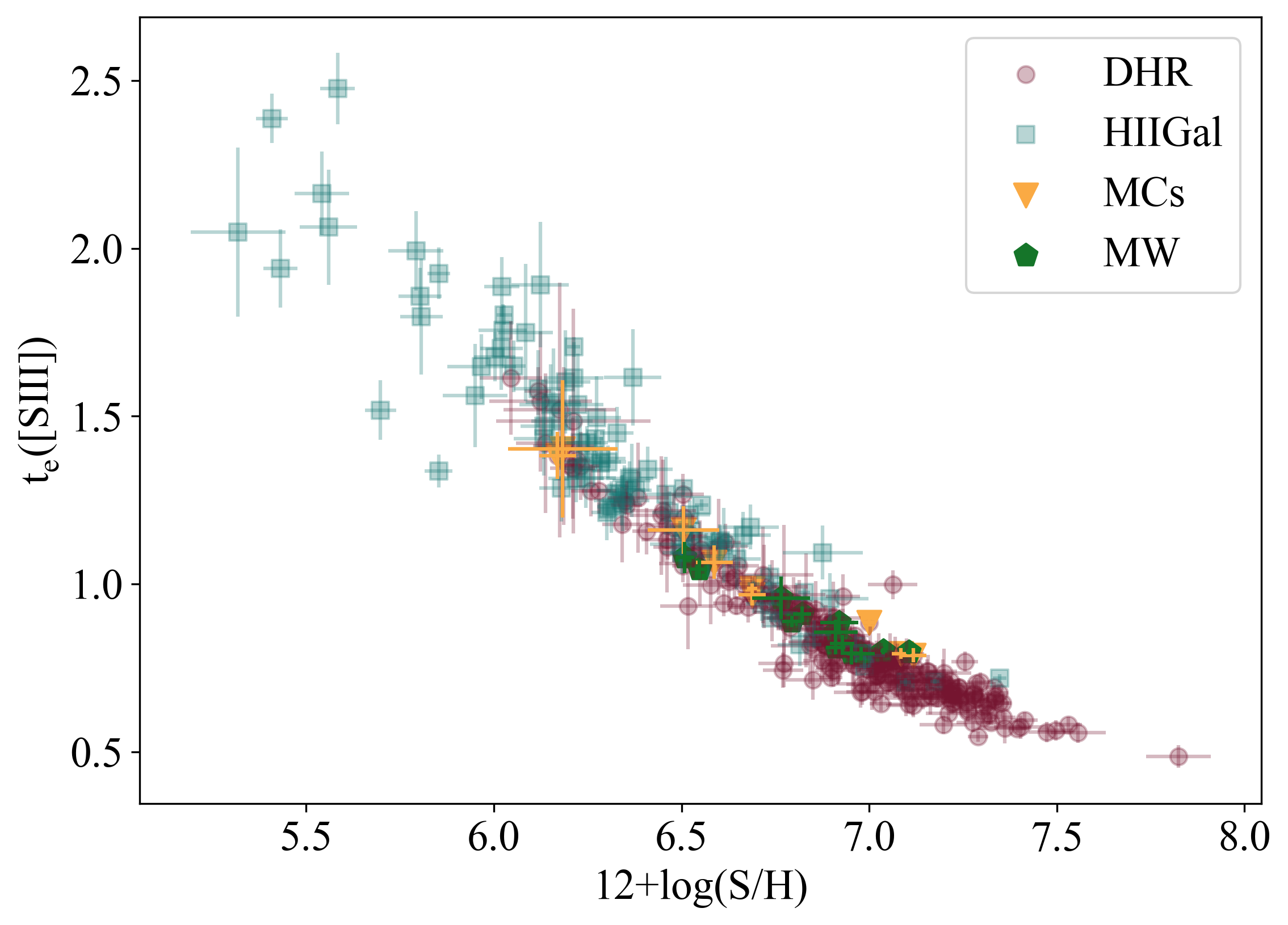}
 \caption{Correlation between  the electron temperature, as measured by T$_e$([SIII]), and the total sulphur abundance: 12 + log(S/H).}
 \label{fig:te-s}
\end{figure}

The electron temperature of the ionised gas, given by T$_e$([SIII]), was shown in Figure \ref{hist_te} above to have different distributions for the two studied sub-samples. The median value of T$_e$ for the disc \HII\ region sample (DHR) is $\sim$ 7900 K while for the \HII\ galaxy sample (\HII Gal) is ~13000 K. Since in ionised regions the cooling is exerted by the emission lines corresponding to the different metals, this mostly reflects the different average metallicities of the objects in the two sub-samples. The effect of sulphur ions as cooling agents is evident by the good correlation shown between the total sulphur abundance and the electron temperature as measured by T$_e$([SIII]) (Figure \ref{fig:te-s}) and this relation is the base of the empirical calibrations presented in Section 4 above. 

\begin{figure}
 \includegraphics[width=\columnwidth]{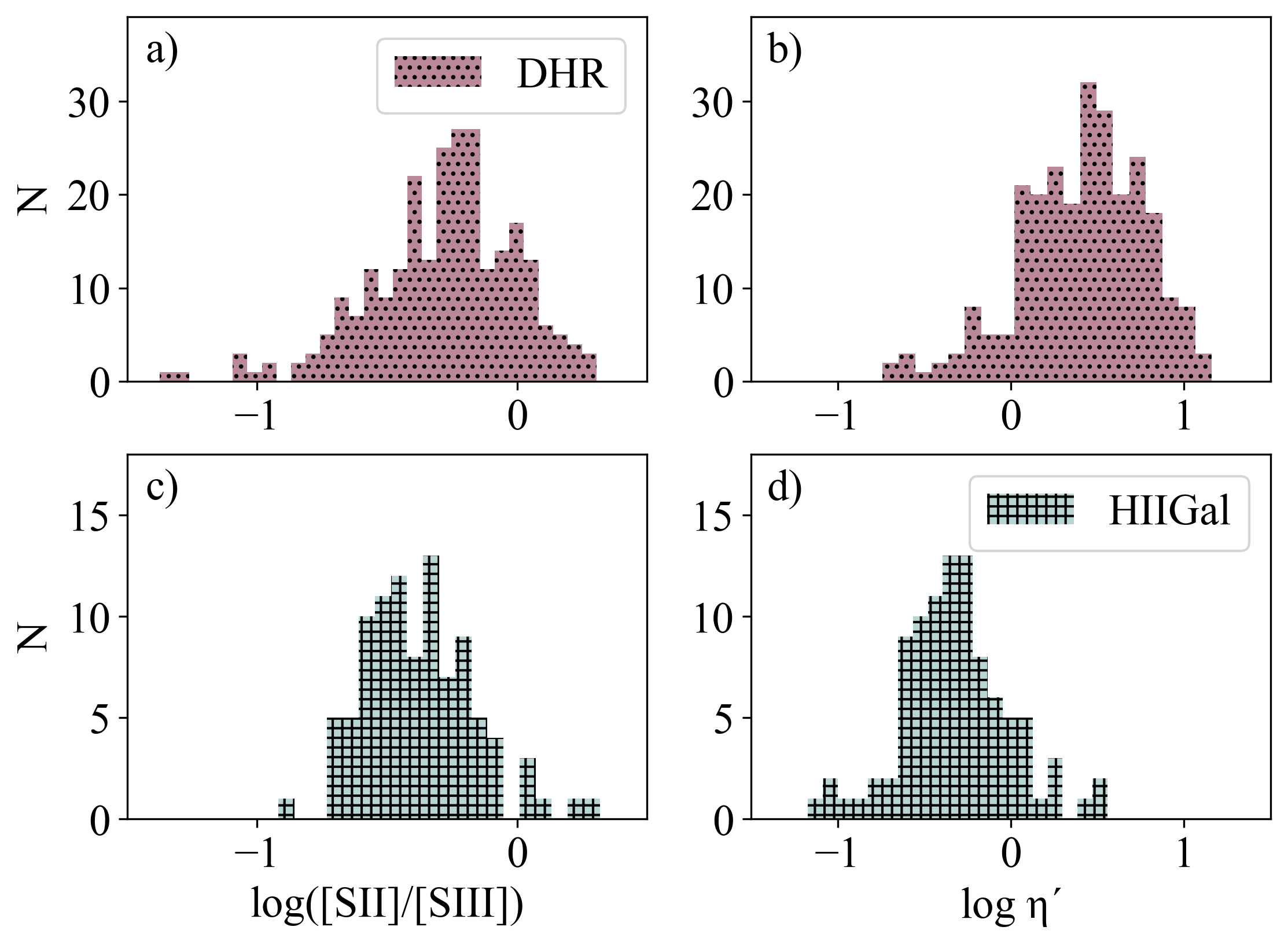}
 \caption{Distribution of the [SII]/[SIII] excitation ratio, taken as a proxy for ionisation parameter(left panels), and the $\eta '$ parameter, taken as a proxy for the ionising temperature (right panels), for objects in sub-sample DHR (upper panels) and \HII\ Gal (lower panels).}
 \label{fig:hist}
\end{figure}

\begin{figure}
 \includegraphics[width=\columnwidth]{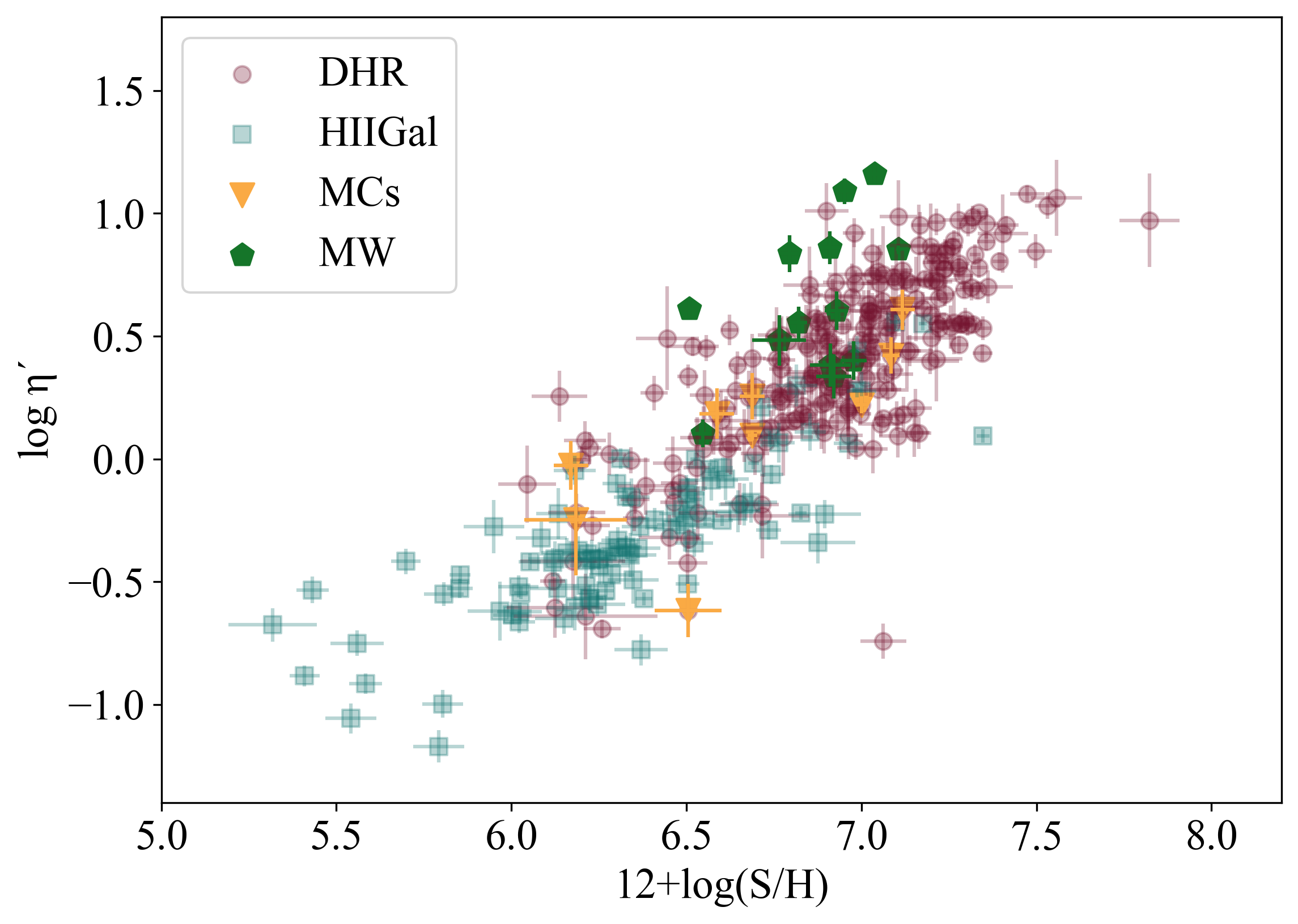}
 \caption{Correlation between the $\eta$' parameter and the total sulphur abundance: 12+log(O/H).}
 \label{fig:eta-S}
\end{figure}

Figure \ref{fig:hist} shows the distribution of ionisation parameter and ionising temperature, as parametrised by the [SII]/[SIII] excitation ratio and $\eta$' respectively, for the two sub-samples. The two of them show rather similar distributions regarding excitation, with median values of log([SII]/[SIII]) of -0.25 for the disc \HII\ region sample (DHR) and -0.38 for the \HII\ galaxy sample (\HII\ Gal). These values translate into ionisation parameters $logu$ of -2.56 and -2.34 according to the calibration given in \citet{1991MNRAS.253..245D}. The situation, however, is very different for the distributions of the values of the $\eta$' parameter which are clearly different for the two sub-samples showing median values of 0.45 for the DHR objects and -0.36 for the \HII\ Gal objects which, according to \citet{1988MNRAS.235..633V}, would correspond to stellar effective temperatures of about 38000 K and 55000 K (the scale being somewhat model dependent) respectively. The similarity between the [SII]/[SIII] distributions can be justified by selection effects since the requirement of the detection of the weak auroral [SIII] line can introduce a certain bias in both sub-samples precluding the selection of lower ionisation parameter \HII\ regions. On the other hand, the different distribution of ionising temperatures seems to correspond to a real effect probably related to the nature of the ionising clusters as shown in Figure \ref{fig:eta-S}.  In fact, \HII\ galaxies are characterised by their low metallicities and young ages, while disc \HII\ regions show a spread of abundances up to the solar values, and maybe beyond, and many of them are more evolved. Both high stellar metallicities and cluster evolution would lead to lower ionising star temperatures. Also, a different behaviour is shown by the objects in the two sub-samples in Figure  \ref{fig:oii/oiii-sii/siii} where diagonal lines would correspond to different values of $\eta$' \citep[see, for example][]{2008MNRAS.383..209H}.

{\begin{figure}
\centering
 \includegraphics[width=\columnwidth]{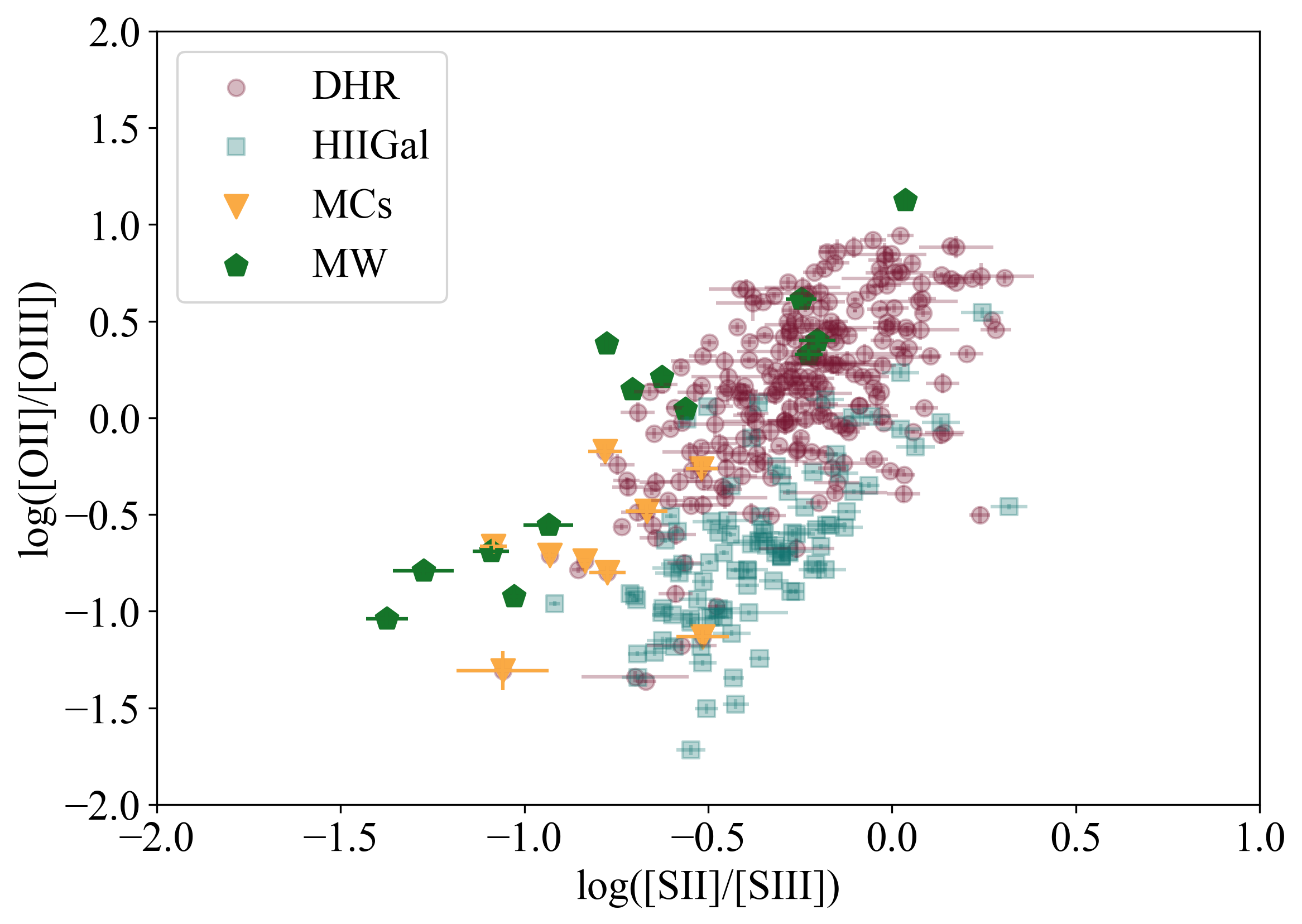}
 \caption{Logarithmic relation between [OII]/[OIII] and [SII]/[SIII] for the two sub-samples. Diagonal lines would correspond to different values of $\eta$'.}
 \label{fig:oii/oiii-sii/siii}
\end{figure}}

\subsection{Abundance results}

\subsubsection{Ionic abundances}

For both sub-samples, S$^{++}$ is found to be the dominant ionisation specie as can be seen in the right panel of Figure \ref{fig:hist_S}, with S$^{++}$/S median values of about 0.7. This fact justifies the assumption of T$_e$([SIII]) as the representative electron temperature for the ionised regions. However, in many cases, and mainly for \HII\ galaxies, corrections regarding the contribution by S$^{3+}$ are required.

Most ICF proposed in the literature are based on the ionisation potential of oxygen ions as compared to those of sulphur with  \citet{1969BOTT....5....3P} first suggesting the use of the ionic ratio $(O^+ + O^{++})/O^+$ as the factor to correct for the lack of observational data on [SIV] emission lines. Later, \citet{1978A&A....66..257S}, based on photoionisation models, proposed the use of the expression: 

\[(S^+ + S^{++})/H^+ = \left[1-\left(1-\frac{O^+}{O}\right)^{\alpha}\right]^{\frac{1}{\alpha}}\]

\noindent to calculate the ICF for sulphur as a function of the O$^+$/O ionic ratio. Different fits to observations by several authors give values of $\alpha$ between 2 and 3 \citep[see, for example,][]{2006A&A...449..193P,2016MNRAS.456.4407D}.

As an alternative to this functional form, in recent works \citep{2018MNRAS.478.5301F}, we have proposed the use of the relation between the ionic ratios of Ar and S in contiguous ionisation stages: $Ar^{+2}/Ar^{+3}$ vs. $S^{2+}/S^{3+}$ which, according to photoionisation models, is found to be linear.

As mentioned in Section 3.1 above, the application of this alternative requires the measurement of the [ArIII] and [ArIV] emission lines at $\lambda\lambda$ 7135 and 4740 \AA\ respectively. Unfortunately, not all published works include data on  this latter line, which limits the application of the method to a certain number of cases and can be circumvented by using the empirical calibration for the ICF by means of the $\eta$' parameter as explained in Section 3.1.

For the \HII\ Gal sample objects showing a significant S$^{3+}$ contribution to the total abundance (ICF > 1.5; 0.18 dex in the derived sulphur abundance) which are about 28\% of the \HII\ Gal sub-sample, this contribution is about 45\% in average (see lower panel d) of Figure \ref{fig:hist_ionratio} and is taken into account by the ICF. The highest ICF are found for objects: J0159+0751, J1205+4551 and J0132+4919, taken from the sample of \emph{compact star-forming galaxies with extremely high [OIII]/[OII] flux ratios} from \citet{2017MNRAS.471..548I} which are probably strong Ly$\alpha$ leaking photons and can be considered rather peculiar. 

On the other hand, most of the disc \HII\ regions (DHR sample), show ICF close to unity with only 17 objects (7\%) of the sub-sample having ICF > 1.2 and 5 objects (2\%) having ICF > 1.5. Therefore,as stated in Section 3.1 it can be assumed that, in the case of this family of objects, no ICF corrections are needed.

\subsubsection{Total abundances}

\begin{figure}
 \includegraphics[width=\columnwidth]{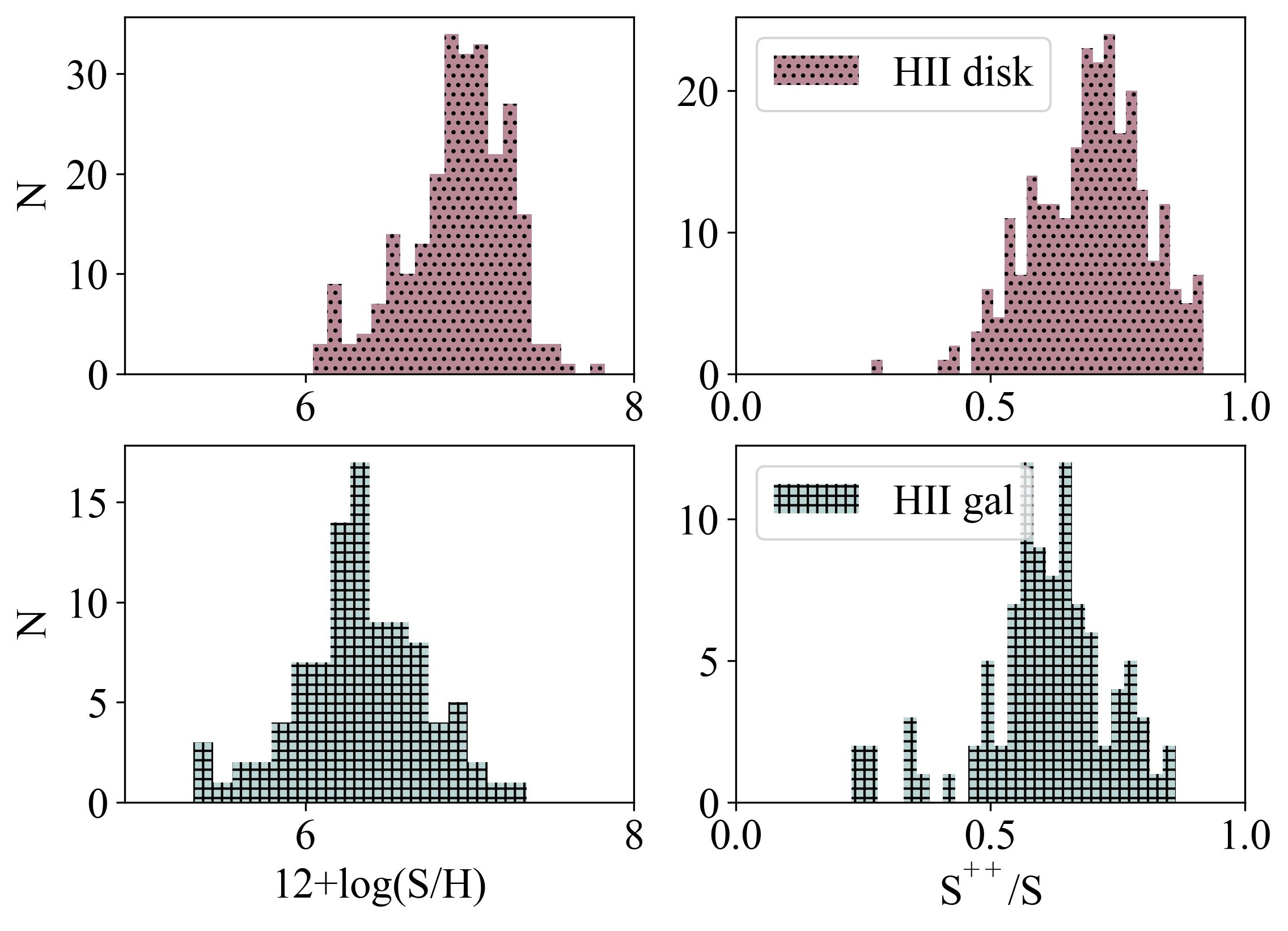}
 \caption{The distribution of the total sulphur abundance (left panels) and the S$^{++}$/S ratio for the two sub-samples: DHR (upper panels) and \HII\ Gal (lower panels). Although the total S/H abundance distribution looks different for the two sub-samples, in both of them $^{++}$ seems to be the dominant ionic specie.}
 \label{fig:hist_S}
\end{figure}

The distribution of the total abundances of sulphur for the objects in the two sub-samples is shown in the left panel of Figure \ref{fig:hist_S}. 

It is evident from the figure that the S/H distributions in both samples are rather different. The median values of 12+log(S/H) are 6.27 in the case of the \HII\ galaxies and 6.92 in the case of the disc \HII\ regions with 25\% of the objects in this sub-sample showing abundances larger than solar, reaching up to 5 times the solar photospheric value \citep[12+log(S/H)$_{\odot}$= 7.12,][]{2009ARA&A..47..481A}. At this high abundances, the [OIII] $\lambda$ 4363 \AA\ auroral line is not detected and hence these objects would had been missing from the metallicity distribution had oxygen  been chosen as abundance tracer. The highest sulphur abundance is found in region 11 of NGC~{\bf 5236} (M~83) \citep{2005AandA...441..981B}. For this object, these authors find T$_e$([SIII])= (4800 $\pm$ 200) K, very close to our derived value of (4858 $\pm$ 338) K and agreeing within the errors. Their corresponding value of the sulphur abundance is 12+log(S/H)= 7.71, also consistent with our derived value of 7.82 $\pm$ 0.08. Therefore, this abundance seems to be well established.

On the other extreme, the lowest  sulphur abundances are found for objects in the \HII\ Gal sub-sample, with the lowest values, of about only  2\% of the solar one, found for SBS0035-052E and IZw18. In these two cases the comparison with the abundances derived by the authors of the articles of reference is difficult. For IZw18SE  \citet{1993ApJ...411..655S}, give 12+log(S/H)= 5.55$\pm$0.06, that should be compared to our derived value of 5.70$\pm$0.04. However, it is not clear which electron temperature is used to derive the S$^{++}$/H$^+$ ionic ratio and no value of T$_e$([SIII]) is given. If T$_e$([OIII]) (17200 K) has been used, this value is higher than the T$_e$([SIII]) derived here by 2000 K which would then lead to lower values of S$^{++}$/H$^+$, and hence lower values of the total S abundance, since this is the dominant specie and the ICF estimated in the two works differ by less that 10\%. For IZw18NW, the value of 12+log(S/H) given by  \citet{1993ApJ...411..655S}, and the one derived in the present work are fully consistent within the errors. This is probably due to the fact that our value of T$_e$([SIII]) (19400 K)  is very close to their value of T$_e$([OIII]) (19600 K). 

The situation is similar for SBS0035-052E. There is a large difference between the value of T$_e$([SIII]) given by \citet{2009AandA...503...61I}, 18704 K, and the one obtained here, 13360 K, for knots 1+2 that translates into larger values of the total sulphur abundance in the latter case by 0.29 dex. Since the oxygen abundance derived by both \citet{2009AandA...503...61I} and the present work is identical (7.28$\pm$0.01), the difference in the sulphur abundance implies a difference in the S/O abundance of about a factor of 2. On the other hand, as in the former case, the results obtained for the other region analysed in this galaxy, knots 4+5, are fully consistent in the two works.

\subsubsection{The sulphur over oxygen abundance}

\begin{figure}
 \includegraphics[width=\columnwidth]{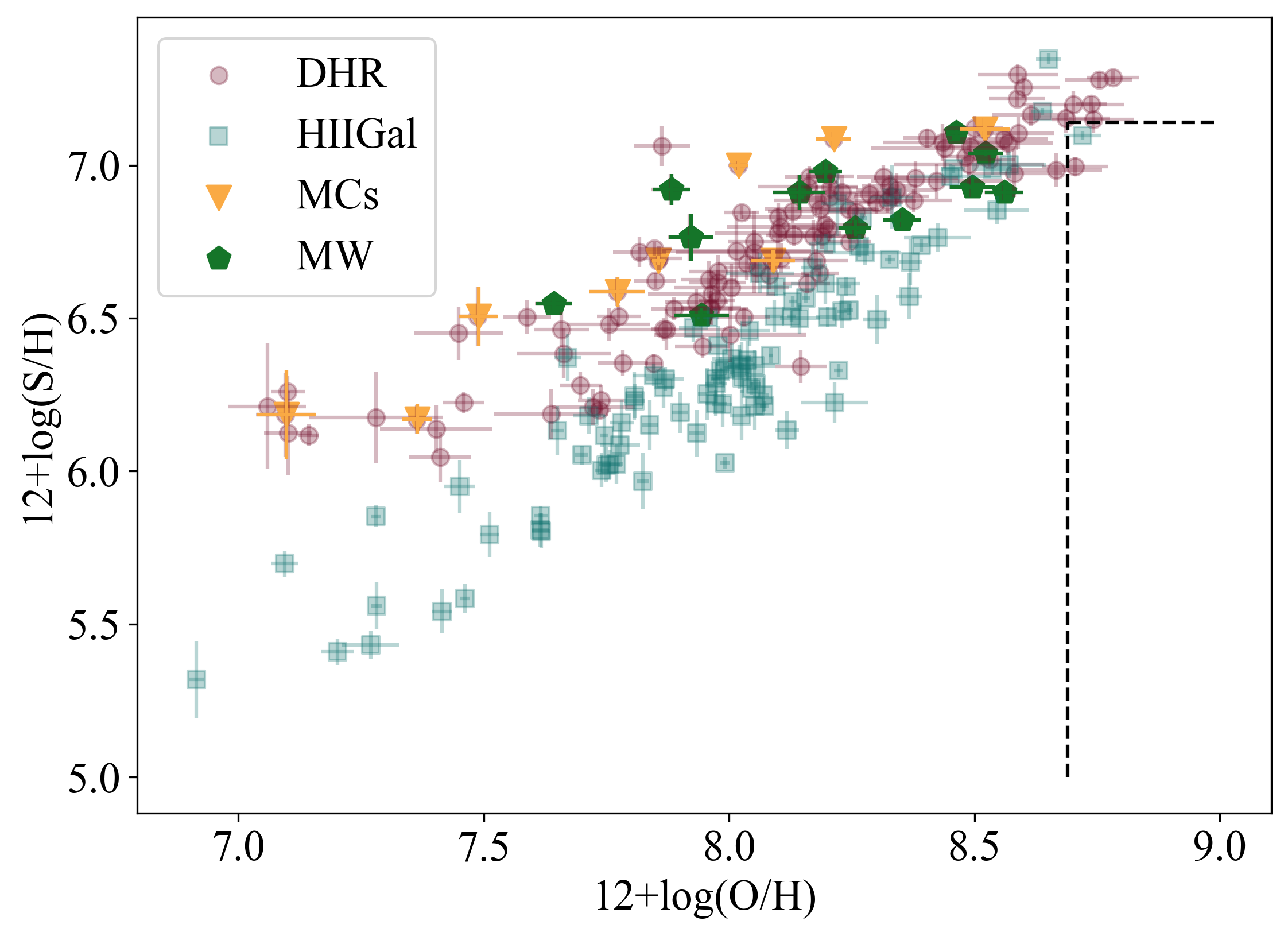}
 \caption{Relation between the sulphur and oxygen abundances of objects in the two sub-samples. The black dashed lines mark the solar photospheric values.}
 \label{fig:O-H-S-H}
\end{figure}

Figure \ref{fig:O-H-S-H} shows the relation between the total abundances of oxygen and sulphur which shows a different behaviour for the the two families of objects studied here, the relation being in both cases almost linear but with different slopes which, in principle, argues against the commonly assumed constancy of the S/O ratio. 

\begin{figure}
 \includegraphics[width=\columnwidth]{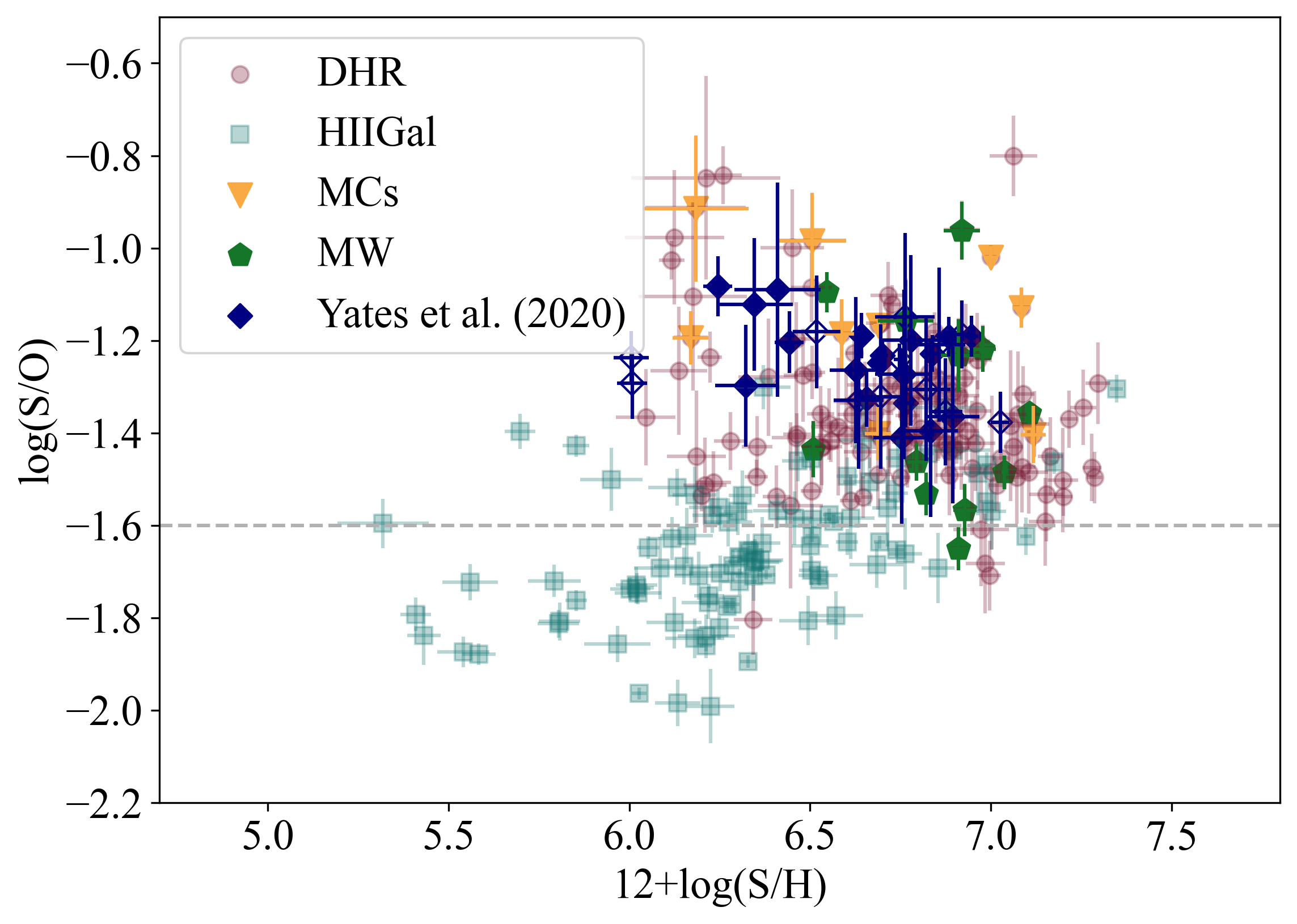}
 \includegraphics[width=\columnwidth]{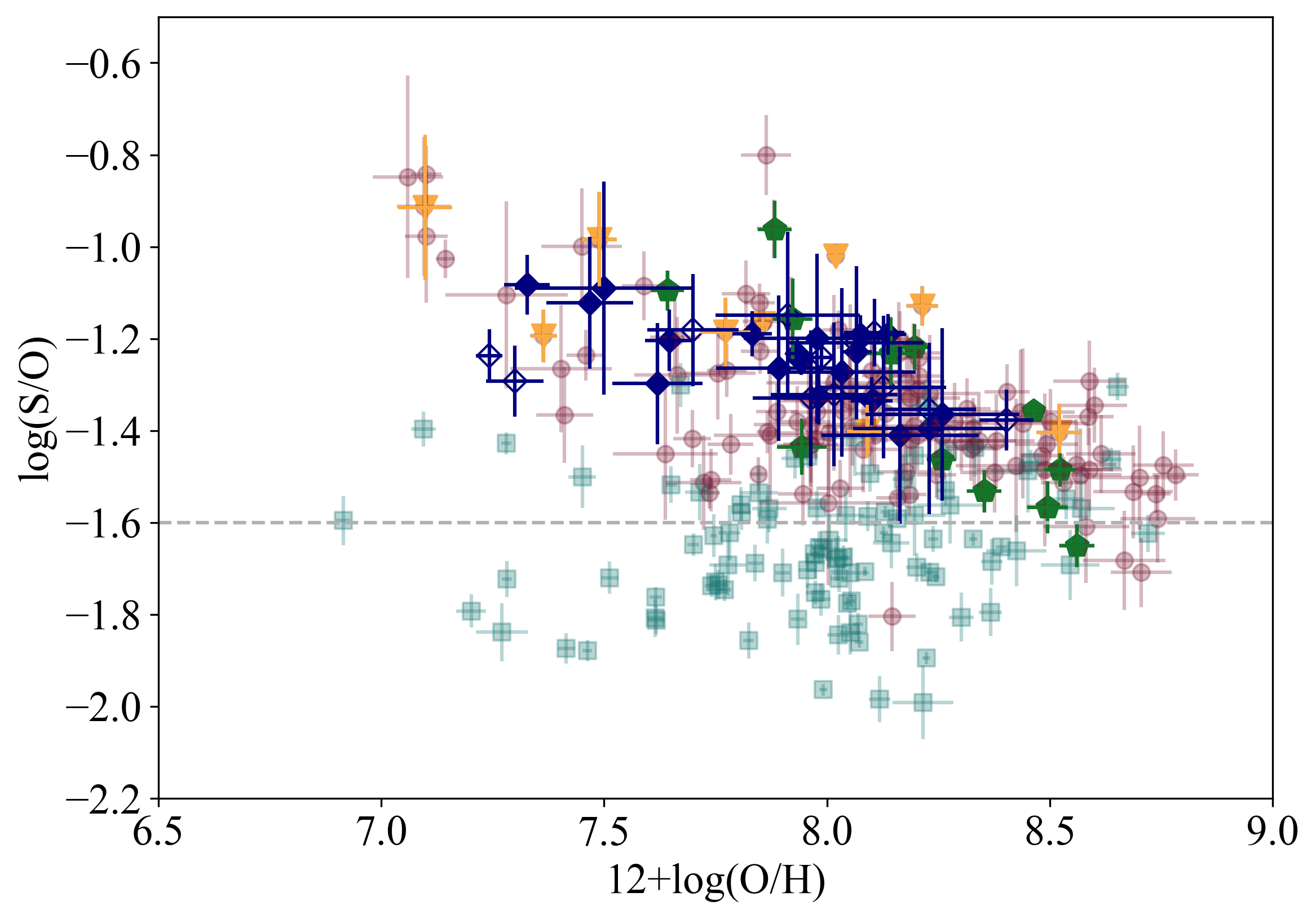}
\caption{S/O relation against the total abundances of sulphur (upper panel) and oxygen (lower panel) for all the objects in the general sample, as labelled. The MANGA survey data by \citep{2020A&A...634A.107Y} are shown as blue diamonds, open for the integrated sample galaxies and filled for selected HII regions within them. The black dashed line in each panel marks the solar S/O ratio.}
 \label{fig:S-O-ratio}
\end{figure}

In fact, this is more clearly seen in Figure \ref{fig:S-O-ratio} where the S/O ratio is shown as a function of the S/H abundance (upper panel) and the O/H abundance (lower panel). There are similarities and differences between these two graphs. In the upper panel a  tendency can clearly be seen for \HII\ galaxies showing an S/O ratio increasing with increasing sulphur abundance; this tendency is however lost in the lower panel where the abundance is traced by oxygen. In both cases, most of the objects show S/O ratios below the solar value which is indicated by the horizontal line. 
\citet{2006A&A...448..955I} invoke oxygen depletion onto grains increasing with metallicity to explain a similar tendency found in \HII\ galaxies for the Ne/O vs O/H relation. This might well apply also in our case being more clearly visible when an undepleted element as sulphur is used as metallicity tracer.

The second tendency, appreciable in both graphs is that exhibited by HII regions. Firstly, they show S/O ratios larger than the solar value and, secondly they show a tendency for lower S/O ratios for higher metallicities as traced by both sulphur and oxygen. This effect has already been noticed by other works \citep{1991MNRAS.253..245D,1997A&A...322...41C,2002A&A...391.1081V,2006MNRAS.367.1139P}, but its reality is difficult to establish due to the different uncertainties involved, mainly in the derivation of the oxygen abundance and the usually large observational errors present up to now, but highly reduced in this work. Some of the uncertainties could also be due to the heterogeneous nature of our general sample, probably affecting more to the DHR sub-sample comprising HII regions of different sizes and geometries. In order to check our results to a more homogeneous HII region sub-sample we have collected the IFU data presented by \citet{2020A&A...634A.107Y} in Tables B.1 for selected HII regions in 12 galaxies from the MANGA survey and B.2 for their global integrated spectra. They are shown in Figure \ref{fig:S-O-ratio} as blue diamonds and seem to follow the same trend as our data.

\begin{figure}
 \includegraphics[width=\columnwidth]{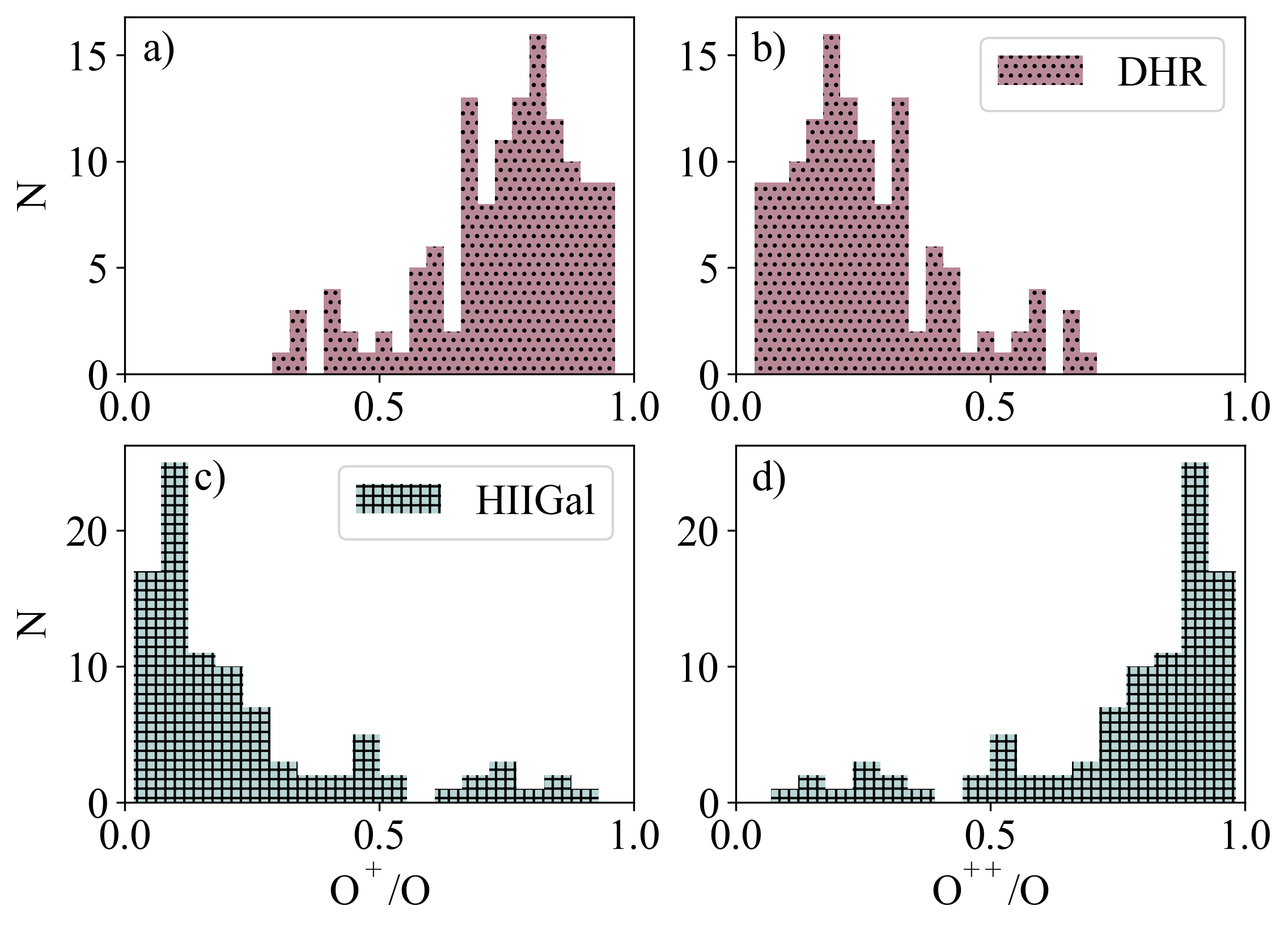}
 
 \vspace{1.0cm}
 
 \includegraphics[width=\columnwidth]{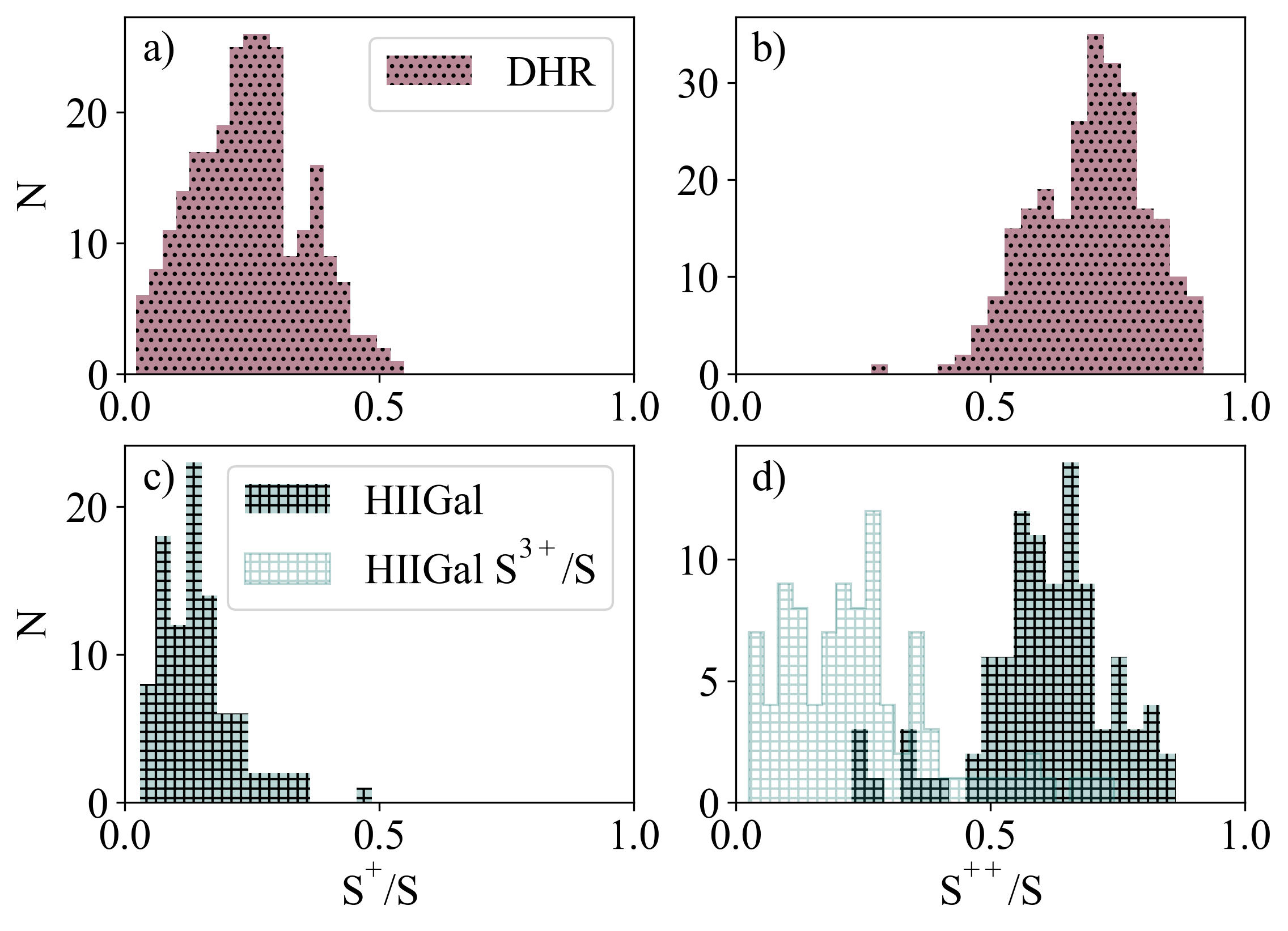}
 \caption{Distribution of the ionic ratios of oxygen (O$^+$ and O$^{++}$ (upper panels) and sulphur (S$^+$, S$^{++}$ and S$^{3+}$ (lower panels) for the two sub-samples as labelled.}
 \label{fig:hist_ionratio}
\end{figure}

\begin{figure}
 \includegraphics[width=\columnwidth]{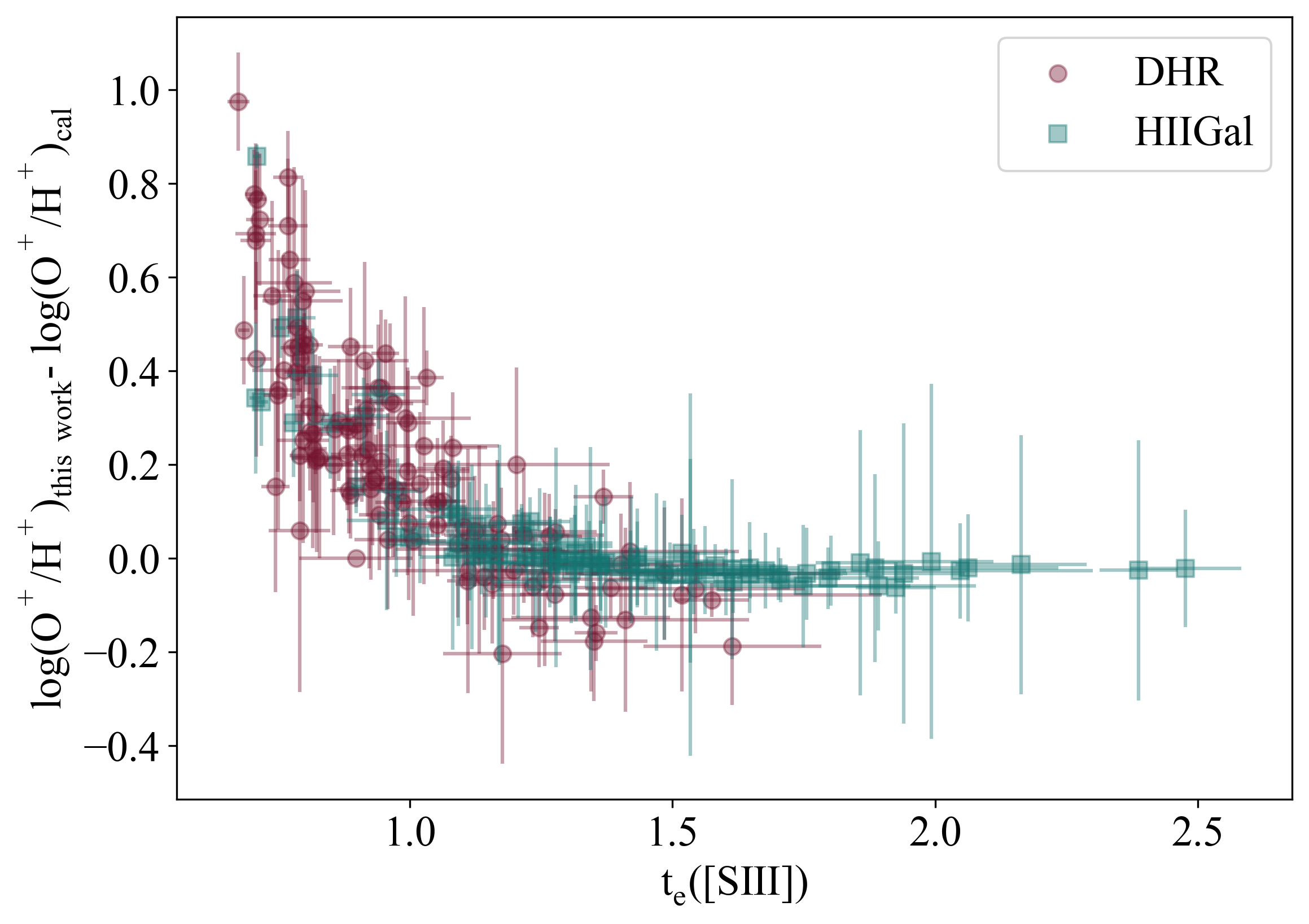}
 \caption{The difference between  O/H abundances derived with the temperature structure assumed in this work, i.e. T$_e$([SIII])=T$_e$([OII])=T$_e$([SII]), and the one commonly assumed, T$_e$([OII])=T$_e$([SII]) with T$_e$([OII]) derived from T$_e$([OIII]), plotted against T$_e$([SIII]).}
 \label{fig:Ab-diff}
\end{figure}

At this point, it is opportune to comment on the uncertainties involved in both oxygen and sulphur abundance derivations (see Section 3). Since the observational errors attached to the strong recombination and collisional line measurements are rather small, these concern essentially the uncertainties in the derivation of the electron temperatures involved, which have different effects depending on the ionisation structure of the nebulae. Figure \ref{fig:hist_ionratio} shows the distribution of the ionic ratios of oxygen (upper panel) and sulphur (lower panel) one and twice ionised. In the case of oxygen only the O$^+$ and O$^{++}$ species are involved, hence mirror distributions are found for the two studied sub-samples. In \HII\ galaxies most of the oxygen is in the form of O$^{++}$, and a large uncertainty in the assumed value of T$_e$([OII]) from the choice of different photoionisation model sequences, as described in Section 3.2, affecting the derivation of the O$^+$/H$^+$ abundance ratio does not translate into large uncertainties in the derivation of the total O/H abundance. However, the opposite can be said about disc \HII\ regions where most of the oxygen is in the form of O$^{+}$, and hence a large uncertainty in the assumed value of T$_e$([OII]) translates into large errors in the derived total O/H abundance. The formal error attached to the derivation of T$_e$([OII]) is the propagation of the error in the derived  T$_e$([OIII]) value and no uncertainty is attached to the choice of the photoionisation models involved or the model calibration obtained. Figure \ref{fig:Ab-diff} shows the difference in the O/H abundances derived with the temperature structure assumed in this work, i.e. T$_e$([SIII])=T$_e$([OII])=T$_e$([SII]), and the one commonly assumed, T$_e$([OII])=T$_e$([SII]) with T$_e$([OII]) derived from T$_e$([OIII]) with the help of Eq. 10, against T$_e$([SIII]). As can be seen in the figure, almost no difference exists for T$_e$([SIII]) $\gtrsim$ 11000 K, but the difference becomes progressively larger as T$_e$([SIII]) decreases, which stems from the fact that Eq. 10 overestimates T$_e$([OII]) for low temperature regions (something recently pointed out also by \citet{2020A&A...634A.107Y}), in fact out of the range of the temperature relation, which leads to an underestimation of the O$^+$/H$^+$ which is the major contributor to the O/H abundance. Therefore, the O/H abundances derived for moderate to high metallicity \HII\ regions using the standard method should be taken with caution.

On the other hand, our assumption of T$_e$([SIII]) = T$_e$([OII]) = T$_e$([SII]) seems to be well justified in view of recent works that find a tight relation between directly derived T$_e$([SIII]) and T$_e$([NII]), this latter one characterising the low ionisation zone \citep{2004ApJ...615..228B,2015ApJ...808...42C,2020ApJ...893...96B}.
Besides, in both \HII\ galaxies and disc \HII\ regions most of the sulphur is in the form of S$^{++}$, and the error in the derived total S abundance just depends on the error in the derivation of T$_e$([SIII]). To this, one should add the corresponding uncertainty introduced by the estimated ICF for the approximately 50\% of the \HII\ galaxies for which there is some contribution by S$^{3+}$ (light green histogram in lower right panel), at any rate amounting, in average, to less than 30\% and included in the error budget. 

If the effect of a decreasing S/O ratio with increasing metallicity in \HII\ regions is real, as it seems, it is expected to be related to stellar nucleosynthesis and chemical evolution. Theoretically, sulphur is mainly synthesised in hydrostatic and explosive burning of O and Si, and stars with masses greater than 10 M$_{\odot}$ are the most likely candidates for its production. The S/O ratio produced is greater for stars with masses in the range 12-20 M$_{\odot}$; then, in principle, variations in the IMF for massive stars could produce variations in the S/O ratio. However, the situation could be more complex if there were a significant contribution from SNIa to the S yield as has lately been found \citep[see for example][]{1999ApJS..125..439I}. Recent work on the the chemical composition of evolved stars (upper giant branch) along the galactic disc (3 to 15 kpc) presenting the trends of elemental abundance ratios, S/O among them, shows very little SNIa contribution to the sulphur enrichment, contrary to theoretical yield results, and a decrease of the S/O ratio with increasing metallicity, as shown by our data, is evident although with a smaller slope \cite{2019ApJ...874..102W}. Obviously more detailed work is needed, mostly in the high-metallicity regime, in order to assess the reality of the trends of S/O with metallicity shown by \HII\ regions.

\subsection{Empirical calibrations}

The observation of the nebular sulphur lines provides an excellent calibrator, S$_{23}$, for empirical abundance determinations, since they are easily observable in objects covering a wide range in overall abundance, the calibration can be done in a purely empirical way without relying on photoionisation models or extrapolations and it remains single valued up to metallicities well beyond the solar abundance.

The calibrations presented in Figure \ref{fig:calibracion} differ only in the application of the ICF correction which affects mostly to the regions showing the highest excitation (lower metallicities). The version shown in the upper panel can be used with confidence for low excitation \HII\ regions which constitute the bulk of the giant extragalactic regions observed over the discs of spirals. 

Although the abundance empirical calibrations or strong line methods assume that ionised regions are essentially characterised by their chemical composition, the degree of ionisation and the ionising temperature, as well as the electron density to a lesser degree, are also involved in the production of the emission line spectrum. Therefore, a good abundance calibrator should not be very dependent on these secondary parameters. Actually this is one of the problems encountered when parameters involving emission lines of oxygen and nitrogen are used \citep[see, for example][]{1991ApJ...380..140M,2002MNRAS.330...69D,2005MNRAS.361.1063P,2007A&A...462..535Y,2013A&A...559A.114M}.

\begin{figure}
\centering
\includegraphics[width=\columnwidth]{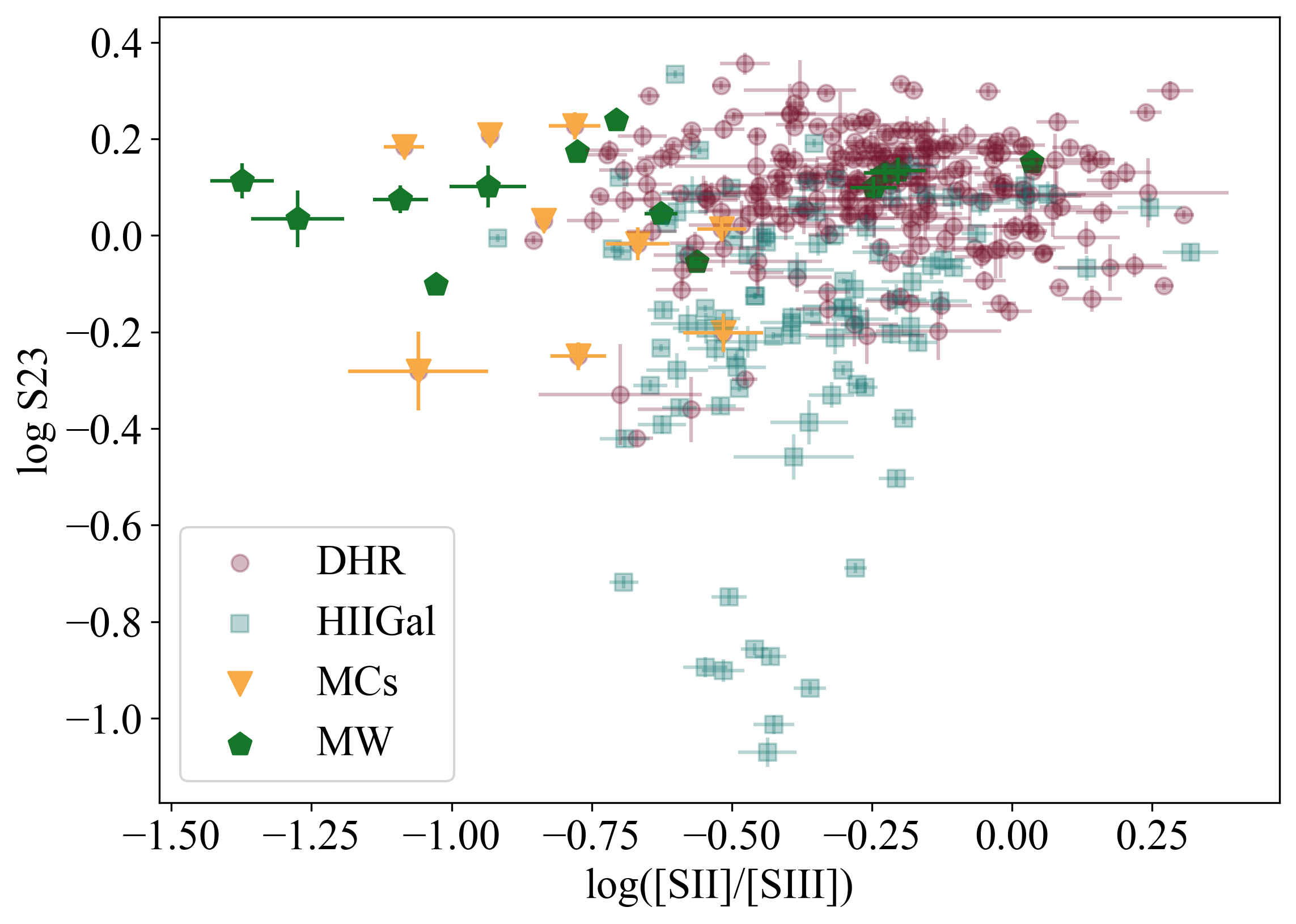}
\includegraphics[width=\columnwidth]{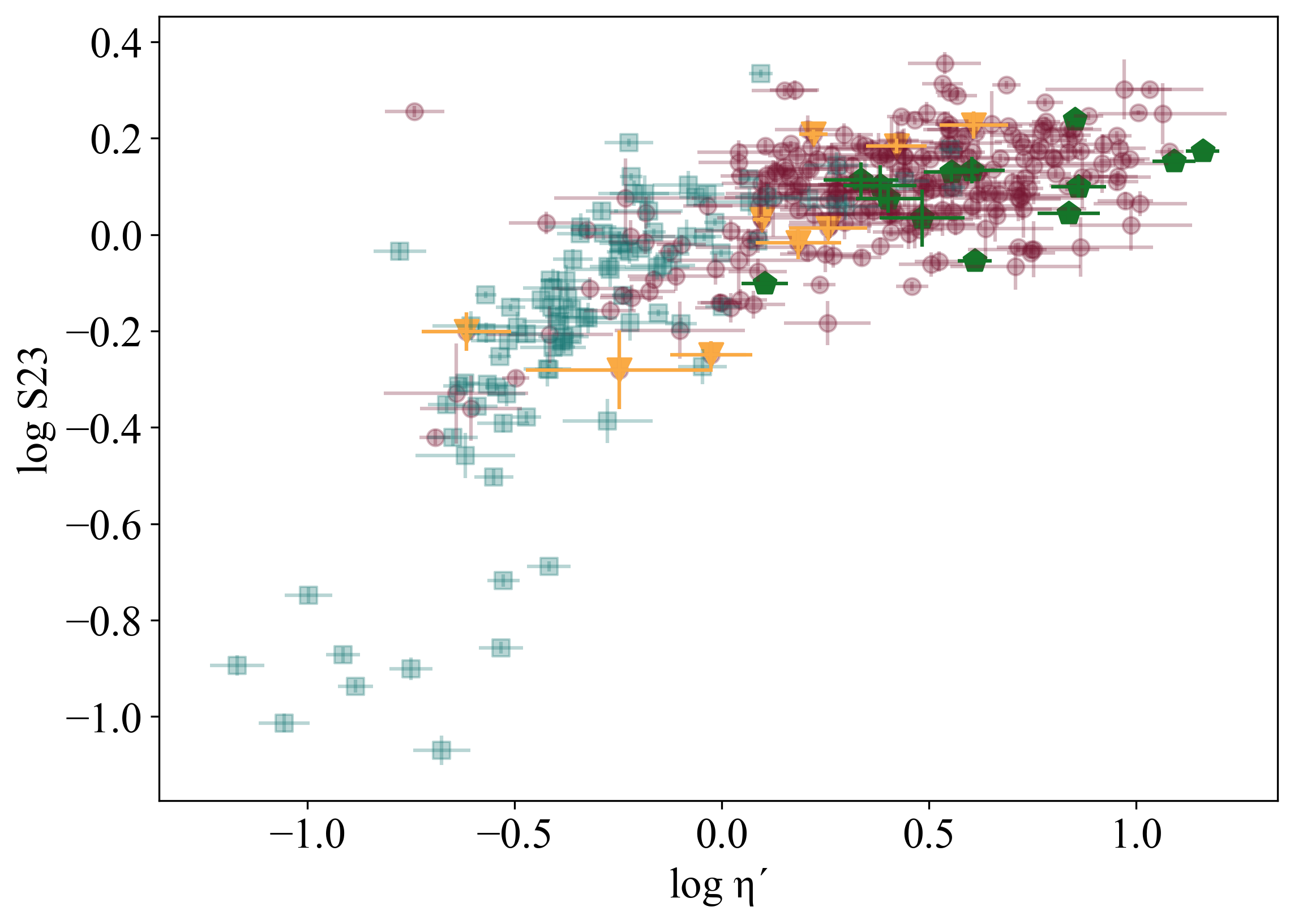}
 \caption{Dependence of the S$_{23}$ from the [SII]/[SIII] excitation ratio, tracing the ionisation parameter, u (upper panel), and the $\eta$' parameter parameter tracing ionising temperature (lower panel).}
 \label{fig:parameter-s23}
\end{figure}

\begin{figure}
\centering
\includegraphics[width=\columnwidth]{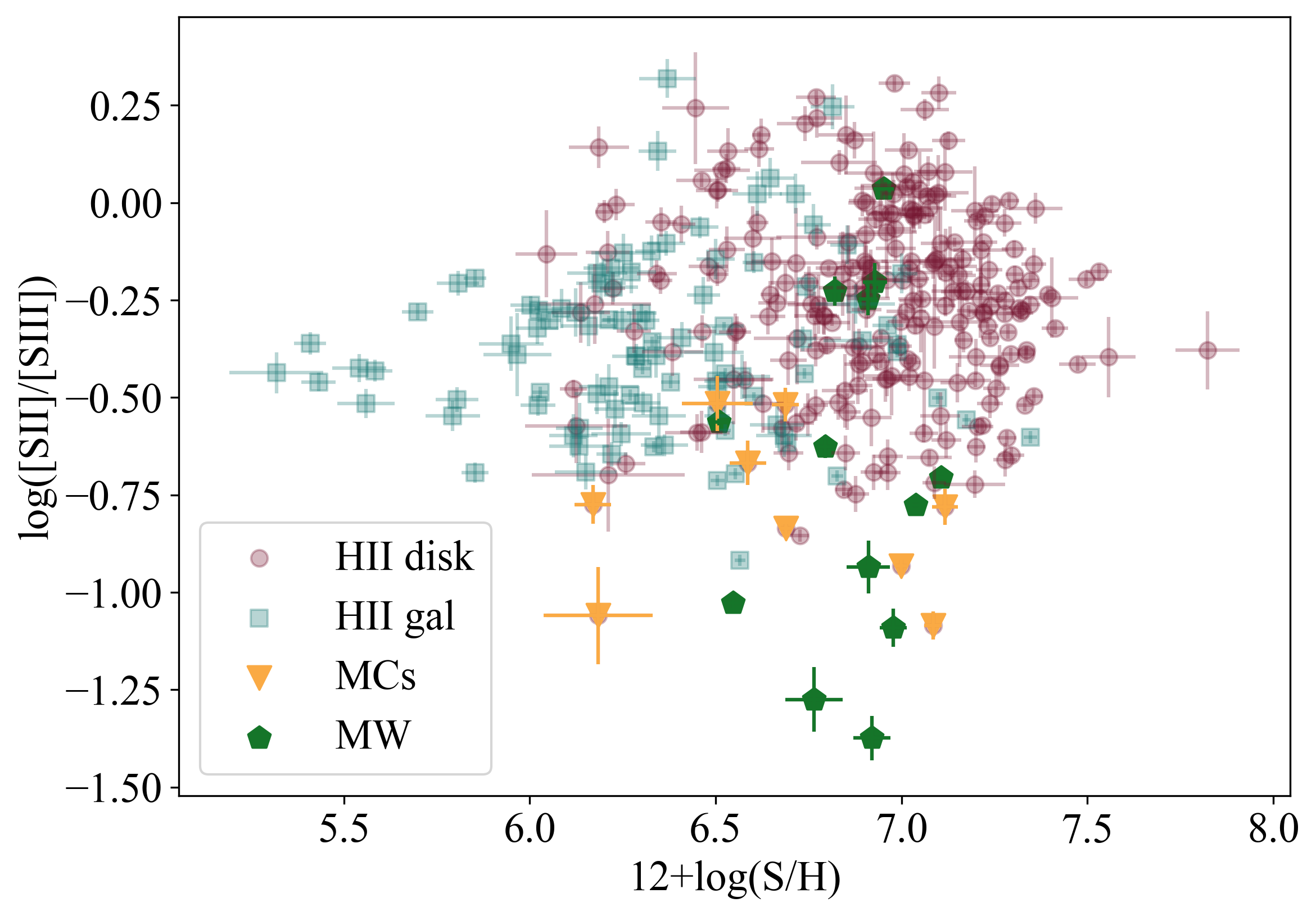}
 \caption{The excitation ratio [SII]/[SIII], taken as a proxy for the ionisation parameter, as a function of total sulphur abundance.}
 \label{fig:S-Total-u}
\end{figure}

The dependence of S$_{23}$ on ionisation parameter and ionisation temperature can be explored using the [SII]/[SIII] line ratio \citep{1991MNRAS.253..245D} and the $\eta$' one \citep{1988MNRAS.235..633V}. The upper panel of Figure \ref{fig:parameter-s23} shows that there is no apparent correlation between ionisation parameter and S$_{23}$, with objects with the lowest metallicities (log(S$_{23}$)< -0.6) showing an almost constant value of the [SII]/[SIII] ratio and the rest of objects showing a wide range of values at a given S$_{23}$. On the other hand, the lower panel of the figure shows a clear relation between this parameter and $\eta$' in the sense that lower metallicity objects, mainly \HII\ galaxies and \HII\ regions in the Magellanic Clouds, show lower values of $\eta$' (higher ionising temperatures) than objects with higher metallicities, mainly disc \HII\ regions, something already reported by \citet{2006A&A...457..477K} in their study of a small sample of \HII\ galaxies for a more restricted parameter range. These relations stem from the fact that a tight correlation exists between $\eta$' and total sulphur abundance (see Figure \ref{fig:eta-S}) while, contrary to what is commonly assumed, no relation seems to exist between ionisation parameter and metallicity as can be seen in Figure \ref{fig:S-Total-u}.

\begin{figure}
\centering
 \includegraphics[width=\columnwidth]{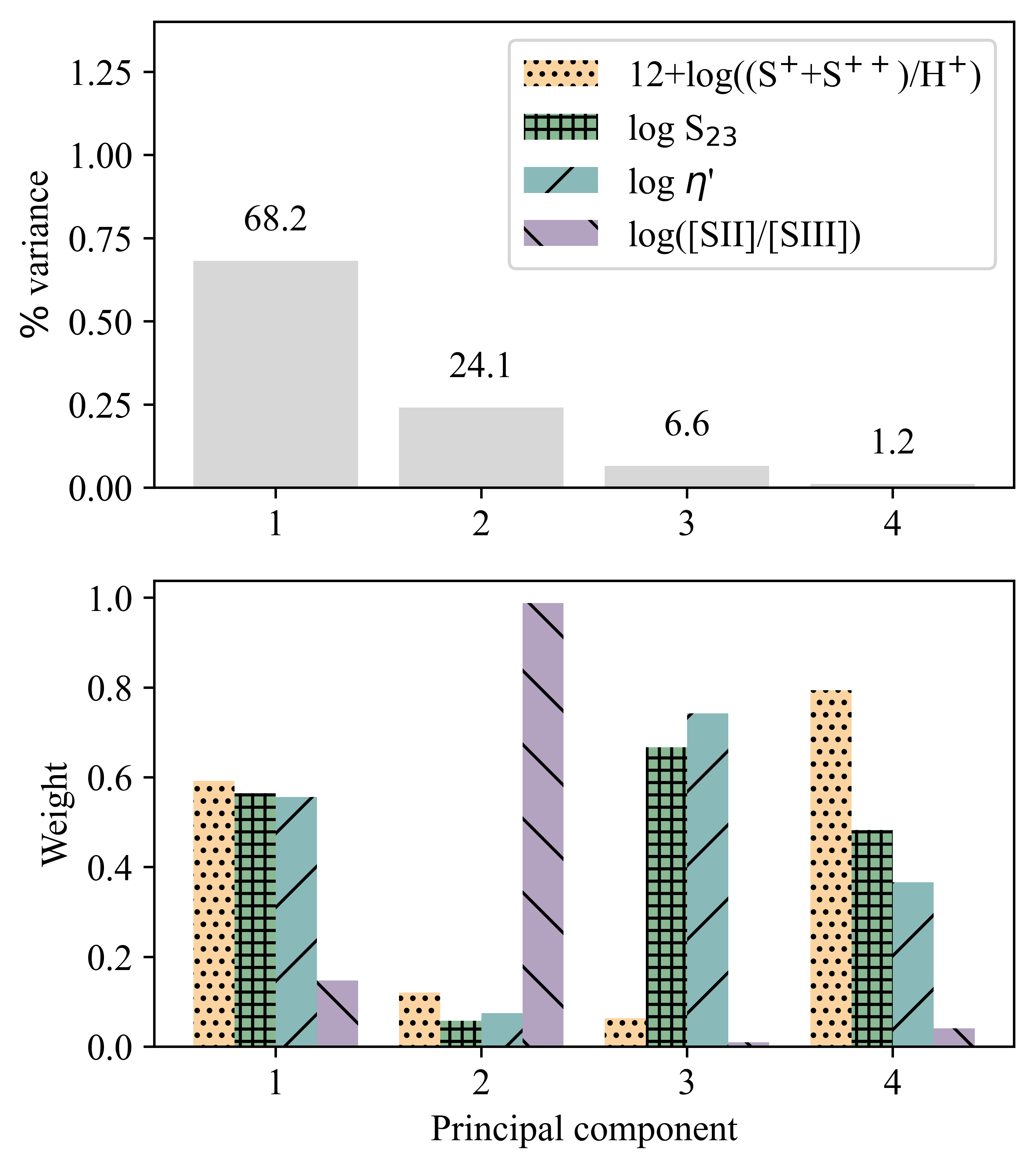}
 \caption{Results of the PCA showing the percentage contribution of each principal component to the total variance shown by the data set (upper panel) and the normalised relative weights of each parameter, as labelled, to each principal component (lower panel).}
 \label{fig:PCA}
\end{figure}

To better understand the relation between the S$_{23}$ parameter and the functional parameters: S/H, u, and $\eta$' we have performed a principal component analysis (PCA). The obtained result clearly confirms the independence of the ionisation parameter, u, from the rest of the variables and hence it has been excluded from further analysis. The relation among the other three parameters involved explains 76 \% of the variance shown by the data set, with very similar weights by each of them (Fig. \ref{fig:PCA}). This tendency is maintained in the analysis of the calibration residuals. According to this, we have made a correction to the derived log(S/H) from the calibration given by eq. \ref{eq_14} to take into account the correlation of its residuals with $\eta$'. This correction is a separate term, a function of log($\eta$'), that should be added to the original calibration and reduces the typical deviations by 14\% for log(S/H) and 10\% for log(S$_{23}$). This correction is negative for log($\eta$') > 0.301 and therefore decreases the derived S/H abundances by a small amount mostly for regions ionised by low temperature sources.

\begin{equation}
f_{cor} [log(\eta ')]= 0.055 - 0.124 \cdot log(\eta ')- 0.195\cdot log(\eta ') ^2
\label{eq_15}    
\end{equation}

\section{Summary and conclusions}

In this work we have explored the use of sulphur to trace chemical abundances as deduced from the observation of ionised star-forming regions, developing a methodology to derive physical properties and abundances of ionised regions using mainly spectroscopy in the red-to-near infrared wavelength range, from 600 to 980 nm.  The method involves the analysis of the emission lines of [SIII] $\lambda$ 6312 \AA , [SII] $\lambda$ 6717,31 \AA\ and [SIII] $\lambda$ 9069,9532 \AA , analogous to the optical [OIII] $\lambda$ 4363 \AA , [OII] $\lambda$ 3227,29 \AA\ and [OIII] $\lambda$ 4959,5007 \AA\ lines.  This observational set up presents several interesting advantages: the [SIII] $\lambda\lambda$ 9069,9532 \AA\ are clearly observable over the full metallicity range and the [SIII]$\lambda$ 6312 \AA\ falls in a spectral region of high instrumental sensitivity and is weakly affected by underlying stellar population absorption. These two facts taken together make it possible to derive the sulphur abundances by direct methods up to over-solar values. Besides, the region in the ionised nebula where  $S^{2+}$ originates  practically covers both the $O^{+}$ and $O^{2+}$ ones and hence the [SIII] electron temperature, T$_e$([SIII]), can be taken as representative of the whole nebula. 

We have compiled a large set of high quality, moderate-to-high spectral resolution, published data that have been divided in two sub-samples: \HII\ regions in spiral and irregular galaxies (DHR sub-sample) and dwarf galaxies dominated, at present, by a strong starburst (\HII Gal sub-sample). The two sub-samples show overlapping ranges of ionisation parameter, however they differ widely in the value of the parameter $\eta$' indicating that objects of sub-sample \HII Gal have hotter ionising stellar populations.

For all the objects, we have derived the abundances of sulphur by direct methods under the assumption of an ionisation structure composed of two zones: an intermediate excitation zone where S$^{++}$ is originated and a low ionisation zone where S$^+$ is formed. For both sub-samples, S$^{2+}$ is found to be the dominant ionisation specie, however, in high ionisation zones, the S$^{3+}$ contribution cannot be neglected. We have used an ICF scheme based on the Ar$^{3+}$/Ar$^{3+}$ ratio which requires the detection and measurement of the [ArIV]$\lambda$ 4740 \AA\ line, finding a good correlation between the sulphur ICF and the parameter $\eta$', which is a proxy for the ionisation temperature. We have applied this correlation to estimate the ICF for those objects without data on the [ArIV] line. Only 28\% of the objects of sub-sample \HII Gal show ICF > 1.5 and hence a significant S$^{3+}$ contribution to the total abundance. On the other hand, most of the objects from sub-sample DHR (disc \HII\ regions) show ICFs close to unity and therefore, it is rather safe to assume that, in the case of this family of objects, no ICF correction needs to be applied.

The S/H abundances distributions in both samples span different ranges with median values of 12+log(S/H) of 6.27 in the case of the \HII Gal sub-sample, and 6.92 in the case of the DHR sub-sample with abundances reaching up to 5 times the solar photospheric value in the most extreme case. At this high abundances, the [OIII] $\lambda$ 4363 \AA\ auroral line is not detected and hence these objects would had been missed from the distribution had oxygen been chosen as abundance tracer. The lowest sulphur abundances are found for objects in the \HII\ Gal sub-sample, with the lowest values, of about only  2\% of the solar one. A good correlation exists between the sulphur abundance and $\eta$' such as low metallicity objects, mostly from the \HII Gal sub-sample, seem to be ionised by stellar clusters of higher effective temperature (lower $\eta$' values).


In order to explore the relation between the sulphur and oxygen abundances, we have derived the latter, also using direct methods, for the objects of the general sample that have data on the [OIII] $\lambda$ 4363 \AA . Most of the objects from the \HII Gal sub-sample show S/O ratios below the solar value and a trend is found for an increased S/O ratio with increasing sulphur abundance, that we hypothesise might be due to a degree of oxygen depletion onto grains increasing with metallicity. On the other hand, the objects in the DHR sub-sample show S/O ratios larger than solar and show a tendency for lower S/O ratios for higher metallicities as traced by both sulphur and oxygen. More detailed work is needed, mostly in the high-metallicity regime, in order to assess the reality of this trend and understand its causes.


Finally, we present a calibration of the S$_{23}$ parameter with the S/H abundance both in the form of (S$^+$+S$^{++}$)/H$^+$, i.e. without correcting for the presence of S$^{S3+}$, and in the form of 12+log(S/H) where this latter contribution is included through the calculated ICF. This calibration is an update of those presented \citep{2000MNRAS.312..130D} and \citep{2006A&A...449..193P} and has been produced using only the highest quality spectroscopic data in order to reduce the observational errors and study better the systematic ones. A reexamination of the calibration shows the S$_{23}$ parameter to be independent of the ionisation parameter, but weakly correlated with the $\eta$' parameter, reflecting some dependence of S$_{23}$ on the temperature of the radiation field. An empirical correction is proposed to account for this effect which applies only to the lowest excitation objects whose abundances are overestimated by a small amount. 



\section*{Acknowledgements}

We would like to thank Dr Elena Terlevich for the careful reading of this manuscript and an anonymous referee for his/her suggestions for improvement.
This work has been supported by Spanish grants from the  former Ministry of Economy, Industry and Competitiveness through the MINECO-FEDER research grant AYA2016-79724-C4-1-P, and the present Ministry of Science and Innovation through the research grant PID2019-107408GB-C42. 
S.Z. acknowledges the support from contract: BES-2017-080509 associated to the first of these grants. 

\section*{Data Availability}
The original data on which this article is based can be found in the references listed in Tables \ref{RefDHR}  and \ref{RefHII} of the paper. The data result of the analysis are given in the online supplementary material.

\bibliographystyle{mnras}
\bibliography{Bibliography}

\newpage
\appendix
\label{Tables}
\section*{APPENDICES}
\section{Data source references for DHII sample}
Complete reference table to published data set for objects in the DHR sub-sample of the present compilation: Table \ref{RefDHR-2}.

\begin{table*}
\centering
    \caption{References to published data sets for objects in the DHR sub-sample of the present compilation.}
 \label{RefDHR-2}
\begin{tabular}{ccc}
\hline
ID  & Telescope+Instrument & Reference\\ \hline

1-2 & INT(2.5m)+IDS  & \citet{1987MNRAS.226...19D}\\ 
3 & INT(2.5m)+IDS  & \citet{1988MNRAS.235..633V}\\ 
4-7 & INT(2.5m)+IDS & \citet{1993MNRAS.260..177P}\\ 
8-12 & WHT(4.2m)+ISIS & \citet{1994ApJ...437..239G}\\
13 & MMT(4.5m, 6.5m)+ Blue Channel and Red Chanell Spectrograp & \citet{1994ApJ...426..123G}\\ 
14-16 & WHT(4.2m)+ISIS & \citet{1995ApJ...439..604G}\\ 
17-24 & INT(2.5m)+IDS and KPNO(2.1m)+ GoldCam CCD & \citet{1997ApJ...489...63G}\\ 
25-26 & WHT(4.2m)+ISIS & \citet{2000MNRAS.318..462D}\\ 
27-30 & WHT(4.2m)+ISIS & \citet{2002MNRAS.329..315C}\\ 
31-45 & MMT(4.5m, 6.5m)+ Blue Channel and Red Chanell Spectrograp & \citet{2003ApJ...591..801K}\\ 
46 & VLT(8.2m) + UVES & \citet{2003ApJ...584..735P}\\ 
47-52 & KeckI(10.0m)+LRIS & \citet{2004ApJ...615..228B}\\ 
53 & VLT(8.0m)+UVES & \citet{2004MNRAS.355..229E}\\ 
54 & VLT(8.0m)+UVES & \citet{2004ApJS..153..501G}\\ 
55-72 & KeckI(10.0m)+LRIS & \citet{2005AandA...441..981B}\\ 
73 & VLT(8.0m)+UVES & \citet{2005MNRAS.362..301G}\\ 
74-76 & VLT(8.0m)+UVES & \citet{2006MNRAS.368..253G}\\ 
77-78 & KeckI(10.0m)+LRIS & \citet{2007ApJ...656..186B}\\ 
79-80 & VLT(8.0m)+UVES & \citet{2007RMxAA..43....3G}\\ 
81-104 & VLT(8.0m)+FORS2 Multi-object & \citet{2009ApJ...700..309B}\\ 
105-109 & KeckI(10.0m)+LRIS & \citet{2012MNRAS.427.1463Z}\\ 
110-148 & LBT(8.4m)+MODS & \citet{2015ApJ...806...16B}\\ 
149-165 & LBT(8.4m)+MODS & \citet{2015ApJ...808...42C}\\ 
166-221 & LBT(8.4m)+MODS & \citet{2016ApJ...830....4C}\\ 
222-226 & VLT(8.2m) + UVES & \citet{2016MNRAS.458.1866T}\\ 
227-231 & WHT(4.2m)+ISIS & \citet{2017AandA...597A..84F}\\ 
232-239 & VLT(8.2m) + UVES & \citet{2017MNRAS.467.3759T}\\ 
240-255 & LBT(8.4m)+MODS & \citet{2020ApJ...893...96B}\\ 

 \hline     
\end{tabular}
\end{table*}

\section{Data source references for HII Gal sample}
Complete reference table to published data sets for objects in the \HII Gal sub-sample of the present compilation: Table \ref{RefHII-2}.

\begin{table*}
\centering
    \caption{References to published data sets for objects in the \HII Gal sub-sample of the present compilation.}
 \label{RefHII-2}
\begin{tabular}{cccc}
\hline
ID & Telescope+Instrument & Reference\\ \hline

257-258 & Large Cass  McDonal(2.7m)+MMT(6.2m) Spectrograph & \citet{1993ApJ...411..655S}\\ 
259 & ISIS+WHT (4.2m) + MMT(6.2m) & \citet{1994ApJ...431..172S}\\ 
260-270 & IDS+INT(2.5m) & \citet{2003MNRAS.346..105P}\\ 
271-273 & ISIS+WHT(4.2m) & \citet{2006MNRAS.372..293H}\\ 
274-277 & UVES+VLT(8.2m) & \citet{2007ApJ...656..168L}\\ 
278-284 & ISIS+WHT(4.2m) & \citet{2008MNRAS.383..209H}\\ 
285-287 & FORS+UVES+VLT(8.2m) & \citet{2009AandA...503...61I}\\ 
288 & TRIPLESPEC+APO Telescope (3.5m)+SDSS & \citet{2009ApJ...703.1984I}\\ 
289 & FORS+UVES+VLT(8.2m) & \citet{2010AandA...517A..90I}\\ 
290-311 & FORS+UVES+VLT(8.2m) & \citet{2011AandA...529A.149G}\\ 
312 & FORS+UVES+VLT(8.2m) & \citet{2011AandA...534A..84G}\\ 
313-315 & DIS + TRIPLESPEC+APO Telescope (3.5m) & \citet{2011ApJ...734...82I}\\ 
316-318 & XSHOOTER+VLT(8.2m) & \citet{2012AandA...541A.115G}\\ 
319-326 & DIS +APO Telescope (3.5m)+MMT(6.2m) Spectrograph & \citet{2012AandA...546A.122I}\\ 
327-335 & UVES+VLT(8.2m) & \citet{2014MNRAS.443..624E}\\ 
336-344 & LRIS+Keck(10m) & \citet{2015AJ....150...71H}\\ 
345-347 & LBT(8.4m)+MODS & \citet{2017MNRAS.471..548I}\\ 
348 & LBT(8.4m)+MODS & \citet{2018MNRAS.473.1956I}\\ 
349-352 & ISIS+WHT(4.2m) & \citet{2018MNRAS.478.5301F}\\

 \hline     
\end{tabular}
\end{table*}
\newpage

\bsp	
\label{lastpage}
\end{document}